\documentclass[journal]{IEEEtran}
\IEEEoverridecommandlockouts
\usepackage{cite}
\usepackage{amsmath,amssymb,amsfonts}
\allowdisplaybreaks[4]
\usepackage{algorithm,algorithmic}
\usepackage{bm,bbm}
\usepackage{booktabs}
\usepackage{graphicx}
\usepackage{textcomp}
\usepackage{xcolor}
\usepackage{subfigure}
\newtheorem{assu}{Assumption}
\newtheorem{pb}{Problem}
\usepackage[normalem]{ulem}
\newtheorem{lemma}{Lemma}
\newtheorem{theorem}{Theorem}

\newtheorem{Rem}{Remark}
\newtheorem{Cor}{Corollary}
\usepackage{cancel}
\def\BibTeX{{\rm B\kern-.05em{\sc i\kern-.025em b}\kern-.08em
    T\kern-.1667em\lower.7ex\hbox{E}\kern-.125emX}}
\allowdisplaybreaks
\begin{document}

\title{Sampling for Remote Estimation of an Ornstein-Uhlenbeck Process through Channel with Unknown Delay Statistics\\}

\author{
	Yuchao~Chen*, Haoyue Tang*~\IEEEmembership{Member,~IEEE}, Jintao~Wang,~\IEEEmembership{Senior Member,~IEEE}, Pengkun Yang, and Leandros Tassiulas,~\IEEEmembership{Fellow,~IEEE}\\
	
	\thanks{
 *Equal Contribution. 
 
 The work of Y. Chen and J. Wang was supported by the Science, Technology and Innovation Commission of Shenzhen Municipality (NO.JSGG20211029095003004) and Tsinghua University-China Mobile Research Institute Joint Innovation Center. The work of P. Yang is supported by NSFC Grant 12101353, Tsinghua University Initiative Scientific Research Program. The work of H. Tang and L. Tassiulas was supported by the
NSF CNS-2112562 AI Institute for Edge Computing Leveraging Next
Generation Networks (Athena) and the ONR N00014-19-1-2566. 
		
	Y. Chen, J. Wang are with Beijing National Research Center for Information Science and Technology (BNRist) and the Department of Electronic Engineering, Tsinghua University, Beijing 100084, China. J. Wang is also with Key Laboratory of Digital TV System of Guangdong Province and Shenzhen City, Research Institute of Tsinghua University in Shenzhen, Shenzhen, China. (e-mail: \{cyc20@mails.; wangjintao@\}tsinghua.edu.cn)
	
	P. Yang is with the Center for Statistical Science, Tsinghua University, Beijing 100084, China. (email: yangpengkun@tsinghua.edu.cn)
	
	H. Tang and L. Tassiulas are with the Department of Electrical Engineering and Institute for Network Science, Yale University, New Haven, CT, USA. (email: \{haoyue.tang;leandros.tassiulas\}@yale.edu)
	 
	\emph{Corresponding author: Haoyue Tang.}}
}

\maketitle

\begin{abstract}
	In this paper, we consider sampling an Ornstein-Uhlenbeck (OU) process through a channel for remote estimation. The goal is to minimize the mean square error (MSE) at the estimator under a sampling frequency constraint when the channel delay statistics is unknown. Sampling for MSE minimization is reformulated into an optimal stopping problem. By revisiting the threshold structure of the optimal stopping policy when the delay statistics is known, we propose an online sampling algorithm to learn the optimum threshold using stochastic approximation algorithm and the virtual queue method. We prove that with probability 1,  the MSE of the proposed online algorithm converges to the minimum MSE  that is achieved when the channel delay statistics is known. The cumulative MSE gap of our proposed algorithm compared with the minimum MSE up to the $(k+1)$-th sample grows with rate at most $\mathcal{O}(\ln k)$. Our proposed online algorithm can satisfy the sampling frequency constraint theoretically. Finally, simulation results are provided to demonstrate the performance of the proposed algorithm.
\end{abstract}

\begin{IEEEkeywords}
    Ornstein-Uhlenbeck process, online learning, stochastic approximation
\end{IEEEkeywords}

\section{Introduction}

With the rapid development of the autonomous vehicles \cite{IoV2021survey} and intelligent machine communications \cite{MMC2017survey}, status update information (e.g., the speed of the vehicles) is becoming a major part in future communication networks \cite{AoIsuevey2021Modiano}. Those status information are delivered to the destination through communication channels, and to guarantee the system safety and efficient control, it is necessary to ensure that the controller has an accurate estimation of the system state.

To measure the information freshness at the destination, the metric, Age of Information (AoI), has been proposed in \cite{AoI2012Kaul}. According to the definition, AoI measures the difference between the current time and the generation time of the latest information received at the destination. Previous work \cite{Lazy2015Roy,Update2017Sun} have shown that AoI minimization is different from the traditional throughput and delay optimization. Specifically in the data generation procedure, a new data sample should be made only when the data stored at the destination is old.
Numerous research have been conducted to minimize the AoI in various networks \cite{AoI2012Kaul,Lazy2015Roy, AoIMultisource2019Roy,NonAoI2019Sun,Update2017Sun,AoIthrou2019Modiano,AoIInter2020Modiano,AoIStoArr2021Modiano}. The average AoI optimization in the queueing system is studied in \cite{AoI2012Kaul,AoIMultisource2019Roy}. Age-optimal scheduling policies in a multi-user wireless network are also investigated in \cite{AoIthrou2019Modiano,AoIInter2020Modiano,AoIStoArr2021Modiano,tang_jsac}.
For minimizing the more general non-linear age function, \cite{NonAoI2019Sun,Update2017Sun} also design the optimal sampling strategies. 

However, when the signal model is known, AoI itself cannot reflect the different signal evolution. As an alternative, a better metric to capture information freshness at the destination is the mean square error (MSE) \cite{Resurvey2019Jog,adapsample2012Rabi,multiwiener2014Nar,ReEst2011Lipsa,GAO201857,MarkovRE2020Jhelum,AoIRe2022Tsai,wiener2020Sun,OU2021Sun}. The sampling strategy to minimize the estimation MSE of a Wiener process is studied in \cite{adapsample2012Rabi,multiwiener2014Nar,wiener2020Sun}. Sampling strategy to minimize an Ornstein-Uhlenbeck (OU) process is investigated in \cite{adapsample2012Rabi,OU2021Sun}. It is revealed that the optimum sampling threshold depends on signal evolution and channel delay statistics. When the channel delay statistics is known, the aforementioned optimum sampling thresholds can be computed numerically by fixed-point iteration \cite{AoIRe2022Tsai} or bi-section search \cite{wiener2020Sun,OU2021Sun}.

When the channel statistics of the communication link is unknown, finding the optimum policy (i.e., the optimum AoI \cite{Update2017Sun} or signal difference threshold \cite{wiener2020Sun,OU2021Sun}) is challenging. Designing an adaptive sampling and transmission strategy under unknown channel statistics for data freshness optimization can be formulated into a sequential decision-making process \cite{AoIlimit2020Baner,agebandit2021Modiano,AoIbandit2022Fatale,Effi2021Bin, online2021Modiano,age2022thy,Tsai2020AoIRT,Tsai2023ToN}. Based on the stochastic multi-armed bandit, \cite{AoIlimit2020Baner,agebandit2021Modiano,AoIbandit2022Fatale} design online channel selection algorithms to minimize average AoI performance for the ON-OFF channel with unknown transition probability. 
For channels with more efficient communication protocols, \cite{DRLAoI2019lsy,RLAoI2020Pappas,RLAoI2021Gunduz} use reinforcement learning to minimize the AoI performance under unknown channel statistics. For communication channels with random delay, \cite{tanginfocom,Tsai2020AoIRT,Tsai2023ToN} apply the stochastic approximation method to design adaptive sampling algorithms to optimize AoI performance. The stochastic approximation method can also be extended to online estimation of the signals with simple evolution model, i.e., the Wiener process \cite{thy2022wiener}.

Notice that the Wiener process is the simplest time-varying signal model, and we are interested in extending the results to handle more general and complex signal models. In this paper, we consider a point-to-point link with a sensor sampling an OU process and transmitting the sampled packet to the destination through a channel with random delay for remote estimation. Our goal is to design an online sampling policy to minimize the average MSE under a frequency constraint when the channel statistics is unknown. The main contributions of the work are listed as follows:

\begin{itemize}
	\item We reformulated the MSE minimum sampling problem under the unknown channel statistics as an optimal stopping problem by providing a novel frame division algorithm that is different from \cite{OU2021Sun}. This novel approach of frame division enables us to propose an online sampling algorithm to learn the optimal threshold adaptively through stochastic approximation and virtual queue method.
 
	\item When there is no sampling frequency constraint, we proved that the expected average MSE of the proposed algorithm can converge to the minimum MSE almost surely. Specifically, we first utilized the property of the OU process to bound the threshold parameter (Lemma \ref{lem:alpha_lb} and Lemma \ref{lem:bound}), and then we proved the cumulative MSE regret grows at the speed of $\mathcal{O}(\ln K)$, where $K$ is the number of samples (Theorem \ref{thm:mmse}) we have taken.
	
	\item When there exists a sampling frequency constraint, by viewing the sampling frequency debt as a virtual queue, we proved that the sampling frequency constraint can be satisfied in the sense that the virtual queue is stable (Theorem \ref{thm:freq}).
\end{itemize}

The rest of the paper is organized as follows. In Section II, we introduce the system model and formulate the MSE minimization problem. In Section III, we reformulate the problem into an optimal stopping optimization and then propose an online sampling algorithm. The theoretical analysis of the proposed algorithm is provided in Section IV. In Section V, we present the simulation results. Finally, conclusions are drawn in Section VI.

\section{Problem Formulation}

\subsection{System Model}\label{SysModel}

\begin{figure}
	\centering
	\includegraphics[width=0.8\columnwidth]{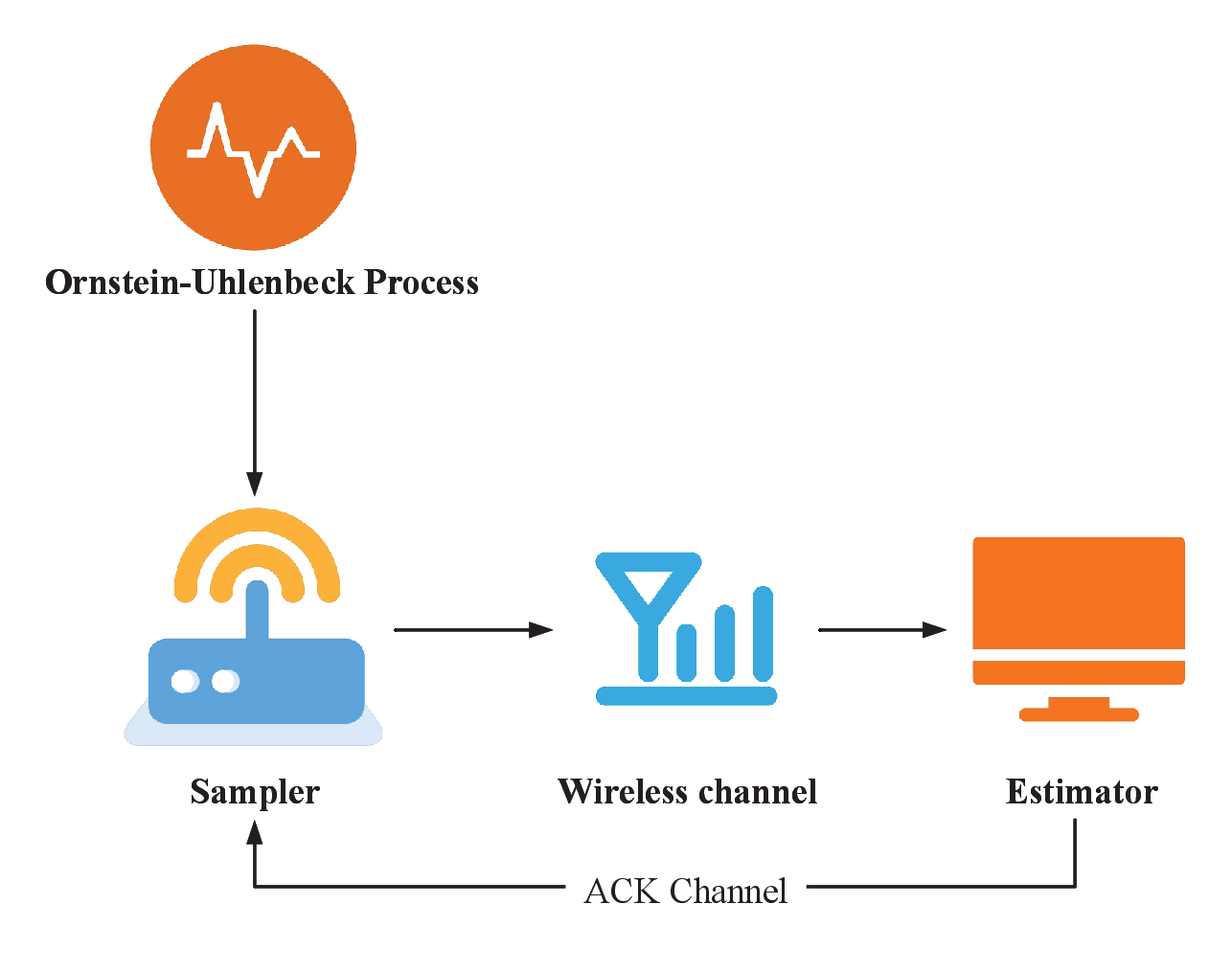}
	\caption{A point-to-point status update system.}
	\label{Fig:sysmodel}
\end{figure}

As depicted in Fig.~\ref{Fig:sysmodel}, we study a status update system similar to \cite{OU2021Sun}, where a sensor observes a time-varying process and sends the sampled data to the remote estimator through a channel. Let $X_t\in\mathbb{R}, \forall t\geq 0$ denote the value of the time-varying process at time $t$. To model these time-varying first-order auto-regressive processes, we assume $X_t$ to be an OU process in this work. This general process is the only nontrivial continuous-time process that is stationary, Gaussian, and Markovian \cite{doob1942brownian}. The OU process evolution parameterized by $\mu,\theta,\sigma \in \mathbb{R}^+$ can be modeled by the following stochastic differential equation (SDE) \cite{doob1942brownian}:
\begin{equation*}
	\text{d}X_t = \theta (\mu-X_t) \text{d}t + \sigma \text{d}W_t,
\end{equation*}
where $W_t$ is a Wiener process.

Suppose the sensor can sample the process at any time 
$t \in \mathbb{R}^{+}$ at his own will. Let $S_{k}$ be the sampling time-stamp of the $k$-th sample. Once sample $k$ is transmitted over the channel, it will experience a random delay $D_k \in [0,\infty)$ to reach the destination. We assume the transmission delay is independent and identically distributed (i.i.d.) following a probability measure $\mathbb{P}_D$.

Due to the interference constraint, only one sample can be transmitted over the channel at one time. Once the transmission of an update finishes, an ACK signal will be sent to the sensor without error immediately. Let $R_k$ be the reception time of the $k$-th sample. Then we can compute $R_k$ iteratively by
\begin{equation}\label{eq:R_k}
	R_k = \max \{S_k, R_{k-1}\} + D_k.
\end{equation}

\subsection{Minimum Mean Squared Error (MMSE) Estimation}
The receiver attempts to estimate the value of $X_t$ based on the received packets and the transmission results before time $t$. Let $i(t)=\max_{k\in \mathbb{N}}\{k|R_k\le t\}$ be the index of the latest received sample at time $t$. The evolution of $X_t$ can be rewritten using the strong Markov property of the OU process \cite[equation (8)]{OU2021Sun} as follows.
\begin{align}
	X_t =& X_{S_{i(t)}} e^{-\theta (t-S_{i(t)})} + \mu \left[1- e^{-\theta (t-S_{i(t)})} \right] \nonumber\\
	&+ \frac{\sigma}{\sqrt{2\theta}}e^{-\theta (t-S_{i(t)})} W_{e^{2\theta (t-S_{i(t)})}-1}. \label{eq:X_t}
\end{align}

Let $\mathcal{H}_t:=\left(\{S_k, D_k, X_{S_k}\}_{k=1}^{i(t)}, t\right)$ be the historical information up to time $t$. Then, the MMSE estimator at the destination is the conditional expectation \cite{books1994sp}:
\begin{equation}\label{eq:hat_Xt}
	\hat{X}_t = \mathbb{E}[X_t|\mathcal{H}_t] = X_{S_{i(t)}} e^{-\theta (t-S_{i(t)})} + \mu \left[1- e^{-\theta (t-S_{i(t)})} \right].
\end{equation}

Combined with \eqref{eq:X_t}, the instant estimation error at time t, denoted by $\Delta_t$ can be computed as
\begin{equation}\label{eq:err}
	\Delta_t = X_t-\hat{X}_t =  \frac{\sigma}{\sqrt{2\theta}}e^{-\theta (t-S_{i(t)})} W_{e^{2\theta (t-S_{i(t)})}-1},
\end{equation}
which can be viewed as an OU process starting at time $t=S_{i(t)}$. 

\begin{figure}
	\centering
	\includegraphics[width=0.8\columnwidth]{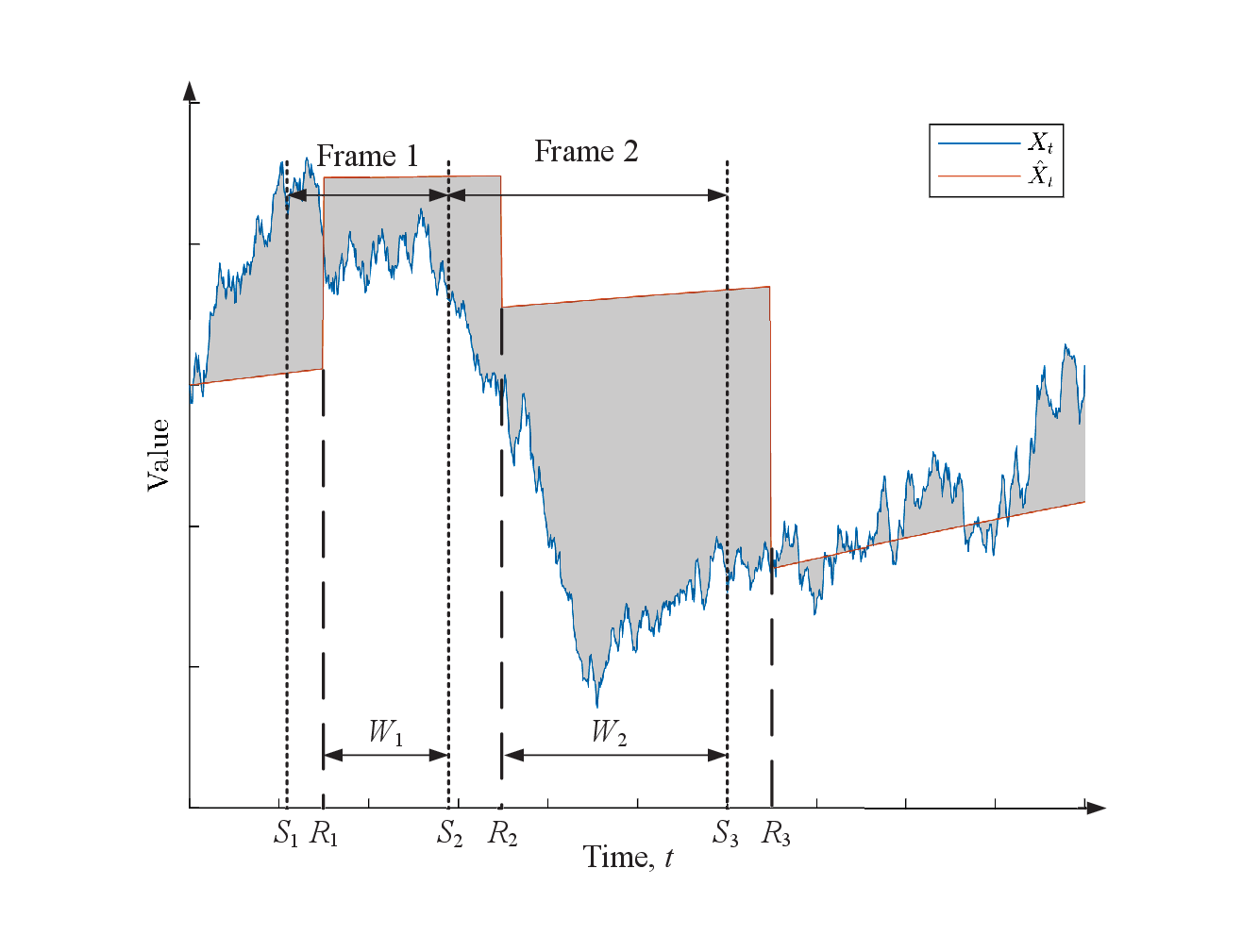}
	\caption{Illustration of the OU process and the estimation error.}
	\label{Fig:ou}
\end{figure}

To better demonstrate the MMSE estimation, we draw Fig.~\ref{Fig:ou} as an example. The blue line is a sample path of an OU process, and the orange line is the MMSE estimator computed by \eqref{eq:hat_Xt}. Then the difference between these two lines, i.e., the shaded area, is the cumulative estimation error between the two samples.

\subsection{Optimization Problem}
The goal of the sampler is to find a sampling policy represented by a series of sampling times, i.e., $\pi:= \{S_1,S_2,\cdots\}$ to minimize the estimation MSE of the OU process at the destination. We assume that the sampler knows the statistical information of the OU process, i.e., parameters $\theta, \mu, \sigma$, while the channel delay statistics $\mathbb{P}_D$ is unknown. Here we focus on the set of causal sampling policies denoted by $\Pi$. The sampling time $S_k$ selected by each policy $\pi \in \Pi$ is determined only by the historical information. No future information can be used for the sampling decision.
Moreover, due to the hardware constraint and energy conservation, the average sampling frequency during the transmission should be below a certain threshold $f_{\text{max}}$. Then, the optimization problem can be formulated as
\begin{pb}[MSE Minimization]{\label{pb:original}}
	\begin{subequations}
		\begin{align}
			\text{mmse} \triangleq &\inf_{\pi \in \Pi} \limsup_{T \to \infty} \frac{1}{T}\mathbb{E}\left[ \int_{0}^T (X_t-\hat{X}_t)^2 \text{d}t \right],\label{eq:primalobj}\\
			&\text{s.t. }\limsup_{T \to \infty} \mathbb{E}\left[\frac{i(T)}{T} \right] \le f_{\text{max}}. \label{eq:samplecons}
		\end{align}
	\end{subequations}
\end{pb}

\section{Problem Resolution}\label{sec:resol}
In this section, we first reformulate the Problem \ref{pb:original} into an optimal stopping problem. Then, an online sampling algorithm is proposed to approach the optimal $\text{mmse}$.

\subsection{Optimal Stopping Problem Reformulation}
Notice that Problem \ref{pb:original} is a constrained continuous-time Markov decision process (MDP) with a continuous state space. It has been proven in \cite[Lemma 6]{OU2021Sun} that it is sub-optimal to take a new sample before the last packet is received by the receiver. In other words, to achieve the optimal $\text{mmse}$, the sampling time-stamp $S_k$ should be larger than $R_{k-1}$. Then \eqref{eq:R_k} can be simplified as $R_k = S_k + D_k$. Let $W_k=S_{k+1}-R_k$ be the waiting time before taking the $(k+1)$-th sample. Then, designing a sampling policy $\pi= \{S_1,S_2,\cdots\}$ is equivalent to choosing a sequence of waiting time $\{W_1,W_2,\cdots\}$. To facilitate further analysis, define frame $k$ to be the time interval between $S_k$ and $S_{k+1}$.
Then, we introduce the following lemma to reformulate the Problem \ref{pb:original} into the packet-level MDP.
\begin{lemma}\label{lemma:RR}
	Define $\mathcal{I}_k = (D_k,\{X_t\}_{t\ge S_k})$ to be the information in frame $k$, and $\Pi_r$ to be the set of stationary sampling policies whose $W_k$ only depends on $\mathcal{I}_k$. Let $D$ be the random delay following distribution $\mathbb{P}_D$. Then
	Problem \ref{pb:original} can be reformulated into the following MDP:
	\begin{pb}[Packet-level MDP Reformulation]{\label{pb:rr}}
		\begin{subequations}
			\begin{align}
				\alpha^\star\triangleq&\sup_{\pi \in \Pi_r}\left(\lim_{K \to \infty} \frac{\sum_{k=1}^K \mathbb{E}[O_{D_k+W_k}^2]}{\sum_{k=1}^K \mathbb{E}\left[D_k + W_k \right]}\right),\label{eq:rr-obj}\\
				&\text{s.t. }\liminf_{K \to \infty} \frac{1}{K}\sum_{k=1}^K \mathbb{E}\left[W_k+D_k \right] \ge \frac{1}{f_{\text{max}}}, \label{eq:rr-cons}
			\end{align}
		\end{subequations}
	where $O_t$ is an OU process with initial state $O_t=0$ and parameter $\mu=0$, which is the solution to the SDE:
    \begin{equation}\label{eq:dOt}
 	\text{d}O_t = -\theta O_t \text{d}t +\sigma \text{d}W_t.
    \end{equation}
    
    Moreover, the optimum value $\alpha^\star$ satisfies:
    \begin{equation}\label{eq:alpha_def}
        \alpha^\star=\left(\frac{\sigma^2}{2\theta} - \text{mmse}\right)\frac{2\theta}{\mathbb{E}[e^{-2\theta D}]}\geq 0. 
    \end{equation}

	\end{pb}
\end{lemma}

The proof of Lemma~\ref{lemma:RR} is provided in Appendix \ref{pf:lemmaRR}.

\begin{assu}\label{assu:D_k}
	The expectation of delay $D_k$ is bounded and known to the transmitter, i.e., 
	\begin{align}
		&0<D_{\text{lb}} \le \overline{D} \triangleq \mathbb{E}_{\mathbb{P}_D}[D_k]\le D_{\text{ub}}<\infty.\label{assu:D_1st}
	\end{align}
\end{assu}

\begin{lemma}\label{lem:alpha_lb}
	Define $\hat{W} = \frac{1}{f_\text{max}} + c$, where $c >0$ is an arbitrary constant. If Assumption \ref{assu:D_k} is satisfied, then we can bound $\alpha^\star$ as
	\begin{equation}
		\alpha_{\text{lb}} \le \alpha^\star \le \alpha_{\text{ub}},
	\end{equation}
	where $\alpha_{\text{lb}}$ and $\alpha_{\text{ub}}$ can be chosen as
	\begin{align}
		&\alpha_{\text{lb}} = \frac{\sigma^2(1- e^{-2\theta\hat{W}})}{2\theta (D_{\text{ub}}+\hat{W})} > 0, \label{eq:alpha_lb}\\
		&\alpha_{\text{ub}} = \sigma^2. \label{eq:alpha_ub}
	\end{align}
\end{lemma}

The proof of Lemma~\ref{lem:alpha_lb} is provided in Appendix \ref{pf:lemmaalpha}. The lower bound is obtained by constructing a feasible and constant sampling policy whose waiting time is always $\hat{W}$ and then using \eqref{eq:rr-obj}. The constant $c$ is introduced to ensure $\hat{W}>0$ when there is no frequency constraint.
The upper bound is obtained by using \eqref{eq:alpha_def} and the fact $\text{mmse} \ge \frac{\sigma^2}{2\theta} \mathbb{E}[1-e^{-2\theta D}]$.


\subsection{Optimal Sampling with Known $\mathbb{P}_D$}\label{subsec:opt}
In the sequel, we will derive the optimum policy $\pi^\star$ that achieves optimal $\text{mmse}$ when $\mathbb{P}_D$ is known. The structure of the optimal policy can help us design the algorithm under unknown channel statistics, and the average MSE obtained by $\pi^\star$ will be used to measure the performance of the proposed online learning algorithm in Subsection~\ref{subsec:alg}

According to \eqref{eq:rr-obj}, the cost obtained by any policy $\pi$ that satisfies the sampling constraint \eqref{eq:rr-cons} is less or equal to $\alpha^\star$. In other words, we have

\begin{equation}\label{eq:gamma^star}
-\lim_{K\rightarrow\infty}\frac{\frac{1}{K}\sum_{k=1}^K\mathbb{E}[O_{D_k+W_k}^2]}{\frac{1}{K}\sum_{k=1}^K\mathbb{E}[D_k+W_k]}\geq -\alpha^\star. 
\end{equation}

Multiplying $\frac{1}{K}\sum_{k=1}^K\mathbb{E}[D_k+W_k]$ on both sides of \eqref{eq:gamma^star} and then adding $\alpha^\star\lim_{K\rightarrow\infty}\frac{1}{K}\mathbb{E}[D_k+W_k]$ on both sides, we are able to solve Problem~\ref{pb:rr} by minimizing the following objective function:
\begin{pb}{\label{pb:frac}}
	\begin{subequations}
		\begin{align}
		\rho^\star\triangleq\inf_{\pi \in \Pi_r} &\limsup_{K \to \infty}\frac{1}{K} \sum_{k=1}^K\big(-\mathbb{E}[O_{D_k+W_k}^2] \nonumber \\ &\qquad\qquad\qquad+\alpha^\star\mathbb{E}[D_k+W_k]\big),\label{eq:fracobj}\\
			\text{s.t. }&\liminf_{K \to \infty} \frac{1}{K}\sum_{k=1}^K \mathbb{E}\left[W_k+D_k \right] \ge \frac{1}{f_{\text{max}}}, \label{eq:fraccons}
		\end{align}
	\end{subequations}
\end{pb}

Similar to Dinkelbach's method \cite{dinkelbach1967nonlinear} for the non-linear fractional programming, we can deduce that the optimal value $\rho^\star$ of Problem \ref{pb:frac} equals 0, and the optimum policy that achieves $\text{mmse}$ in Problem \ref{pb:original} and $\rho^\star$ in Problem \ref{pb:frac} are identical. Therefore, we proceed to solve Problem \ref{pb:frac} using the Lagrange multiplier approach. Let $\lambda\geq0$ be the Lagrange multiplier of the sampling frequency constraint \eqref{eq:fraccons}, the Lagrange function for Problem~\ref{pb:frac} is as follows:
\begin{align}
	\mathcal{L}(\pi,\lambda) =& \limsup_{K \to \infty} \frac{1}{K}\sum_{k=1}^K\big(-\mathbb{E}[O_{D_k+W_k}^2]\nonumber\\
 &\hspace{2cm}+ \left(\alpha^\star-\lambda\right) \mathbb{E}\left[D_k + W_k \right] + \lambda \frac{1}{f_{\text{max}}}\big).\label{eq:lagrange}
\end{align}

Notice that the transmission delay $D_k$ is i.i.d., and $O_t$ is an OU process starting at time $t=0$. Then for fixed $\lambda$, selecting the optimum waiting time $W_k$ to minimize \eqref{eq:lagrange} becomes a per-sample optimal stopping problem by finding the optimum stop time $w$ to minimize the following expectation:
\begin{align}
\min_w \mathbb{E}\left[-O_{D_k+w}^2+(\alpha^\star-\lambda)w|O_{D_k},D_k\right].\label{eq:W_k^*}
\end{align}


For simplicity, let $V_w=O_{D_k+w}$ be the value of the OU process at time $D_k+w$ and $V_0=O_{D_k}$ by definition. Then problem \eqref{eq:W_k^*} is one instance of the following optimal stopping problem when $\beta= \alpha^\star - \lambda$:
\begin{align}
    \sup_{\tau} \mathbb{E}_{v_0}\left[V_\tau^2 - \beta \tau \right], \label{eq:H(v)}
\end{align}
where $\mathbb{E}_{v_0}$ is the conditional expectation given $V_0=v_0$. The optimum policy to \eqref{eq:H(v)} is obtained in the following Lemma:

\begin{lemma}\label{lemma:optstop}
	If $0 < \beta \le \sigma^2$, then the solution to minimize \eqref{eq:H(v)} has a threshold property, i.e.,
	\begin{equation}\label{eq:W_k}
		W_k=w(O_{D_k};\beta):= \inf\{t\ge0:|O_{D_k+t}|\ge v(\beta)\},
	\end{equation}
	where
	\begin{equation}\label{eq:v(beta)}
		v(\beta) = \frac{\sigma}{\sqrt{\theta}} G^{-1} \left(\frac{\sigma^2}{ \beta} \right),
	\end{equation}
	and $G^{-1}(\cdot)$ is the inverse function of
	\begin{equation}
		G(x) = \frac{e^{x^2}}{x}\int_0^x e^{-t^2} \text{d}t,~x\in [0,\infty).\label{eq:thresGdef}
	\end{equation}
\end{lemma}

The proof of Lemma~\ref{lemma:optstop} is provided in Appendix \ref{pf:lemma:optstop}.

Since \cite[Theorem 6]{OU2021Sun} has proven the strong duality of Problem \ref{pb:frac}, i.e., $\rho^\star=\max_\lambda\min\mathcal{L}(\pi, \lambda)$. For notational simplicity, let $o(\beta)$ and $l(\beta)$ denote the expected estimation error and frame length by using threshold $\beta$, i.e., 
\begin{subequations}
    \begin{align}
    o(\beta):=&\mathbb{E}[O_{D+w(O_D;\beta)}^2]\label{eq:odef}\\
    l(\beta):=&\mathbb{E}[D+w(O_D;\beta)]\label{eq:ldef}. 
\end{align}
\end{subequations}
by substituting $O_{D_k+w}$ with $(X_{R_k+w}-\hat{X}_{R_k+w})$ in equation \eqref{eq:W_k},  the optimal sampling time $S_{k+1}=R_k+W_k$ to Problem~\ref{pb:frac} is as follows:
\begin{lemma}\label{lem:beta}
	\cite[Theorem 2 Restated]{OU2021Sun} The optimal solution to Problem~\ref{pb:original} is:
	\begin{equation*}
		S_{k+1} = \inf \{t\ge R_k:|X_t-\hat{X}_t|\ge v(\alpha^\star-\lambda^\star) \},
	\end{equation*}
	where $v(\cdot)$ is defined in \eqref{eq:v(beta)}, $\lambda^\star=\arg\sup_{\lambda}\mathcal{L}(\pi, \lambda)$ is the dual optimizer, and $\alpha^\star$ is the solution to the following equation: 
	\begin{align}\label{eq:opt_b}
            0=&g_{\lambda^\star}(\alpha):=o(\alpha-\lambda^\star)-\alpha l(\alpha-\lambda^\star),
	\end{align}
 where we recall that $o(\beta)=\mathbb{E}[O_{D+w(O_D;\beta)}^2]=\mathbb{E}[(X_{S_{k+1}}-\hat{X}_{S_{k+1}})^2]$ is the expected squared estimation error by using threshold $\beta$, and $l(\beta)=\mathbb{E}[D+w(O_D;\beta)]$ is the expected framelength. 
\end{lemma}

\begin{Rem}
    If the frequency constraint is inactive, then according to the complementary slackness, we have $\lambda^\star=0$, and the threshold becomes $v(\alpha^\star)$. Otherwise, the optimal $\alpha^\star - \lambda^\star < \alpha^\star$. Then according to \eqref{eq:v(beta)}, the sampling threshold is larger than $v(\alpha^\star)$ to satisfy the sampling frequency constraint.
\end{Rem}

\begin{Rem}
	In \cite[Theorem 2]{OU2021Sun}, the optimum sampling threshold to minimize the MSE is
	\begin{equation}\label{eq:v(beta')}
		v(\beta') = \frac{\sigma}{\sqrt{\theta}} G^{-1} \left(\frac{\text{mse}_\infty - \text{mse}_D} {\text{mse}_\infty - \beta'} \right),
	\end{equation}
	where
	\begin{subequations}
		\begin{align}
			&\text{mse}_\infty = \mathbb{E}[O_\infty^2] = \frac{\sigma^2}{2\theta};\label{eq:mse_inf}\\
			&\text{mse}_D = \mathbb{E}[O_{D_k}^2] = \frac{\sigma^2}{2\theta} \mathbb{E}[1-e^{-2\theta D}].\label{eq:mse_D}
		\end{align}
	\end{subequations}
	
	The optimum sampling threshold is taken when $\beta'=\text{mmse}+\lambda'$, i.e., 
	\begin{align}
		v(\beta') &= \frac{\sigma}{\sqrt{\theta}} G^{-1} \left(\frac{\sigma^2} {\left(\frac{\sigma^2}{2\theta} - \text{mmse}\right)\frac{2\theta}{\mathbb{E}[e^{-2\theta D}]} - \lambda' \frac{2\theta}{\mathbb{E}[e^{-2\theta D}]}} \right) \nonumber \\
		&\overset{(a)}{=} \frac{\sigma}{\sqrt{\theta}} G^{-1}\left(\frac{\sigma^2} {\alpha^\star - \lambda' \frac{2\theta}{\mathbb{E}[e^{-2\theta D}]}} \right),\label{eq:rem1term1}
	\end{align}
	where (a) holds by \eqref{eq:alpha_def}. Comparing \eqref{eq:rem1term1} with \eqref{eq:v(beta)}, we find the conclusions coincide.
\end{Rem}

\subsection{Online Algorithm}\label{subsec:alg}
Notice that the optimal sampling in Section \ref{subsec:opt} is determined by $\alpha^\star - \lambda^\star$ through equation \eqref{eq:v(beta)}. However, when the channel statistics $\mathbb{P}_D$ is unknown, $\alpha^\star$ and $\lambda^\star$ are unknown, making direct computation of $v(\alpha^\star - \lambda^\star)$ impossible. To overcome the challenge, we propose an online learning algorithm to approximate these two parameters $\alpha^\star$ and $\lambda^\star$ respectively. 

Notice that $\alpha^\star$ is the solution to equation \eqref{eq:opt_b} when $\lambda=\lambda^\star$. This motivates us to approximate $\alpha^\star$ using the Robbins-Monro algorithm \cite{robbins1951stochastic} for stochastic approximation. For $\lambda^\star$, we construct a virtual queue $U_k$ to record the cumulative sampling constraint violation up to frame $k$.

\begin{algorithm}
	\caption{Online Learning Sampling Algorithm} \label{alg:online}
	\begin{algorithmic}[1]
		\STATE \textbf{Parameters}: $V$.
		\STATE \textbf{Initialization}: 
  $\alpha_1=0,~ U_1 = 0$.
		\FOR{$k=1,2,\cdots,K$}
		\STATE Set $\lambda_k = \frac{1}{V} U_k$.
		\STATE According to the last sampling generation time $S_k$ and delay $D_k$, choose the waiting time $W_k$ as
		\begin{align*}
                &W_k \\
                =& \inf \{w\ge 0:|X_{R_k+w}-\hat{X}_{R_k+w}|\ge v((\alpha_k-\lambda_k)^+)\}.
		\end{align*}
		\STATE Update $\alpha_k$:
		\begin{equation*}
			\alpha_{k+1} = (\alpha_k + \eta_k (O_{L_k}^2-\alpha_k L_k))_{\alpha_{\text{lb}}}^{{\alpha_{\text{ub}}}},
		\end{equation*}
		where
		\begin{align}
			O_{L_k} &= X_{S_{k+1}}-\hat{X}_{S_{k+1}},\label{eq:Q_k} \\
			L_k &= D_k + W_k. \label{eq:L_k}
		\end{align}
		\STATE Update $U_k$:
		\begin{equation*}
			U_{k+1} = \left(U_k + \frac{1}{f_{\text{max}}} - L_k \right)^+.
		\end{equation*}
		\ENDFOR
	\end{algorithmic}
\end{algorithm}

As concluded in Algorithm \ref{alg:online}, the proposed algorithm consists of two parts: sampling (step 5) and updating (step 6 and 7). For the sampling step, the algorithm uses the current estimation $\alpha_k$ and $\lambda_k$ to compute the threshold, i.e.,
\begin{equation}\label{eq:alg_Wk}
	W_k = \inf \{w\ge 0:|X_{R_k+w}-\hat{X}_{R_k+w}|\ge v((\alpha_k-\lambda_k)^+)\},
\end{equation}
where $(\cdot)^+ = \max\{\cdot,0\}$. After sample $(k+1)$ is taken at time $R_k+W_k$, we can compute the instant estimation error $O_{L_k}:=X_{S_{k+1}}-\hat{X}_{S_{k+1}}$ and the frame length $L_k:=D_k + W_k$. According to \eqref{eq:err}, $O_{L_k}$ is an instance of $O_{D+w(O_D;\alpha - \lambda)}$ when $\lambda=\lambda_k$ and $\alpha=\alpha_k$.

We then update $\alpha_{k+1}$ according to the Robbins-Monro algorithm:
\begin{equation}\label{eq:alpha_k}
	\alpha_{k+1} = (\alpha_k+\eta_k(O_{L_k}^2 -\alpha_k L_k))_{\alpha_{\text{lb}}}^{{\alpha_{\text{ub}}}},
\end{equation}
where $(x)_a^b$ is the projection of $x$ onto the interval $[a,b]$; $\alpha_{\text{lb}}$ and $\alpha_{\text{ub}}$ are the lower and upper bound of $\alpha^\star$ defined in \eqref{eq:alpha_lb} and \eqref{eq:alpha_ub}; $\eta_k$ is the step size, which can be chosen as
\begin{align*}
	\eta_k= \begin{cases}
		\frac{1}{2D_\text{lb}}, & k=1; \\
		\frac{1}{(k+2)D_\text{lb}}, & k \ge 2.
	\end{cases}
\end{align*}

For estimating $\lambda^\star$, we construct a virtual queue $U_k$ which evolves as
\begin{equation*}
	U_{k+1} = \left(U_k + \frac{1}{f_{\text{max}}} - L_k \right)^+.
\end{equation*}

Then $\lambda_k = \frac{U_k}{V}$, where $V > 0$ is the hyper-parameter. Notice that $\frac{1}{f_\text{max}} - L_k$ is the violation of sampling constraint in frame $k$. Therefore $U_k$ can be interpreted as the cumulative violation up to frame $k$. The Algorithm \ref{alg:online} attempts to stabilize $U_k$ to satisfy the sampling frequency constraint.

%
%

\begin{Rem}
	In \eqref{eq:alg_Wk}, we choose $(\alpha_k-\lambda_k)^+$ to ensure the positive input for $v(\cdot)$. We should also avoid the estimation $\alpha_k-\lambda_k$ to be zero, which will make the threshold $v$ to be infinite. This requires the algorithm cannot choose $V$ to be too small. Also in practice one can set an arbitrarily small positive value $\eta > 0$ as a lower bound for $\alpha_k-\lambda_k$ to avoid the infinite threshold.
\end{Rem}

\section{Theoretical Analysis}
In this section, we analyze the convergence and optimality of Algorithm \ref{alg:online}.

\begin{assu}\label{assu:D_k^2}
	The second moment of delay $D_k$ is bounded, i.e., \footnote{The assumptions is presented here mainly for theoretical analysis. In fact the proposed algorithm discussed in Section \ref{subsec:alg} does not need the assumption.}
	\begin{subequations}
		\begin{align}
			&0<M_{\text{lb}}\le \mathbb{E}_{\mathbb{P}_D}[D_k^2]\le M_{\text{ub}}<\infty \label{assu:D_2nd}. 
		\end{align}
	\end{subequations}
\end{assu}

First, we assume that there is no sampling frequency constraint, i.e., $f_\text{max}= \infty$ and thus $\lambda = 0$. Finally, we will prove that in general case $f_\text{max} < \infty$, Algorithm \ref{alg:online} will still satisfy the constraint.

\begin{theorem}\label{thm:mmseas}
    The time average MSE $\frac{\int_0^{S_{k+1}}(X_t-\hat{X}_t)^2\text{d}t}{S_{k+1}}$ of the proposed online learning algorithm converges to mmse with probability 1, i.e., 
    \begin{equation}
        \frac{\int_0^{S_{k+1}}(X_t-\hat{X}_t)^2\text{d}t}{S_{k+1}}\overset{\text{a.s.}}{=} \text{mmse}. 
    \end{equation}
\end{theorem}

\begin{theorem}\label{thm:mmse}
    Let $\mathcal{R}_k:=\mathbb{E}\left[\int_0^{S_{k+1}}(X_t-\hat{X}_t)^2\text{d}t\right]-\text{mmse}\cdot \mathbb{E}[S_{k+1}]$ denote the expected cumulative MSE regret up to the $(k+1)$-th sample. We can upper bound $\mathcal{R}_k$ as follows:
 \begin{align}
 \mathcal{R}_{k}\leq \max_{\alpha\in[\alpha_{\text{lb}}, \alpha_{\text{ub}}]}|R_1'(v(\alpha))v'(\alpha)|\frac{\mathbb{E}[e^{-2\theta D}]}{2\theta}\frac{C}{D_{\text{lb}}^2}\ln k,
 \end{align}
 where $C$ is a constant independent of $k$ and is defined \eqref{eq:Cdef}. 
\end{theorem}

The proof of Theorem~\ref{thm:mmseas} and Theorem~\ref{thm:mmse} are provided in Appendix~\ref{pf:thmmmseas} and Appendix~\ref{pf:thmmmse}, respectively.



Now we consider the sampling frequency constraint. Here we assume that the constraint is feasible, i.e.,
\begin{assu}\label{assu:eps}
	There exists a constant $\epsilon>0$, and a stationary sampling policy $\pi_\epsilon$ satisfies
	\begin{equation}
		\mathbb{E} \left[D_k + W_k^\epsilon \right] \ge \frac{1}{f_\text{max}} + \epsilon,
	\end{equation}
	where the expectation is taken over the channel statistics and the policy $\pi_\epsilon$.
\end{assu}

\begin{theorem}\label{thm:freq}
	Under Algorithm \ref{alg:online}, the sampling frequency constraint can be satisfied, i.e.,
	\begin{equation}
		\lim_{K \to \infty} \inf \mathbb{E}\left[\frac{1}{K} \sum_{k=1}^K (D_k + W_k) \right] \ge \frac{1}{f_\text{max}}.
	\end{equation}
\end{theorem}

The proof of Theorem~\ref{thm:freq} is provided in Appendix \ref{pf:thmfreq}.

\section{Simulation Results}
In this section, we provide some simulation results to demonstrate the performance of our proposed algorithm. The parameters of the monitored OU process are $\sigma = 1,\theta = 0.2$, and $\mu = 3$. The channel delay follows the log-normal distribution with $\mu_D = \sigma_D = 1$. The expected MSE is computed by taking the average of 100 simulation runs for $K=10^4$ packet transmission frames. 

\subsection{Without A Sampling Frequency Constraint}
First, we consider the case with no frequency constraint, i.e., $f_\text{max} = \infty$. We compare the MSE performance using the following policies:
\begin{itemize}
	\item \textbf{Zero-Wait Policy $\pi_{\rm{zw}}$}: take a new sample immediately after the reception of the ACK of the last sample, i.e., $W_k$ = 0.
	
	\item \textbf{Signal-Aware MSE Optimum Policy $\pi^\star$}: signal aware MSE optimum policy when $\mathbb{P}_D$ is known \cite{OU2021Sun}.
	
	\item \textbf{Signal-Agnostic AoI Minimum Policy $\pi_{\rm{AoI}}$}: signal agnostic sampling policy for AoI minimization \cite{Update2017Sun}.
	
	\item \textbf{Proposed Online Policy $\pi_{\rm{online}}$}: described in Algorithm \ref{alg:online}.
\end{itemize}

\begin{figure}
	\centering
	\includegraphics[width=0.8\columnwidth]{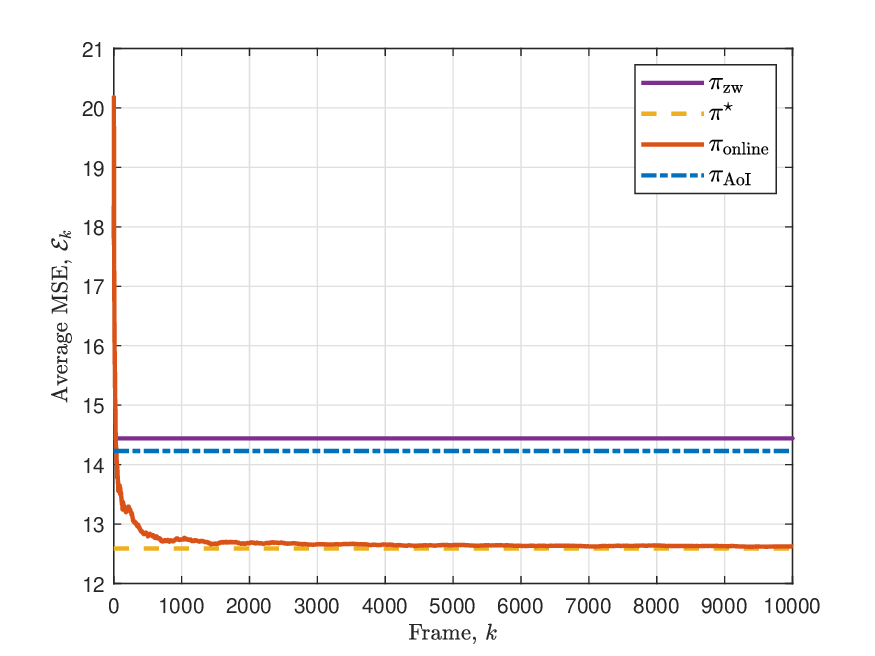}
	\caption{MSE performance with no frequency constraint.}
	\label{Figs:nofreq}
\end{figure}

The estimation performance is depicted in Fig.~\ref{Figs:nofreq}. From Fig.~\ref{Figs:nofreq}, we can verify that the expected MSE performance of the proposed policy $\pi_{\rm{online}}$ converges to the optimum policy $\pi^\star$, and achieves a smaller MSE performance compared with the signal-agnostic AoI minimum sampling and zero-wait policy. Previous work \cite{OU2021Sun} has shown that the zero-wait policy is far from optimality when the channel delay is heavy tail. For the AoI optimal policy, while \cite{wiener2020Sun} reveals the relationship between average AoI and estimation error for the Wiener process, it is sub-optimal for MSE optimization of the OU process, even worse than the zero-wait policy.

\begin{figure}
	\centering
	\includegraphics[width=0.8\columnwidth]{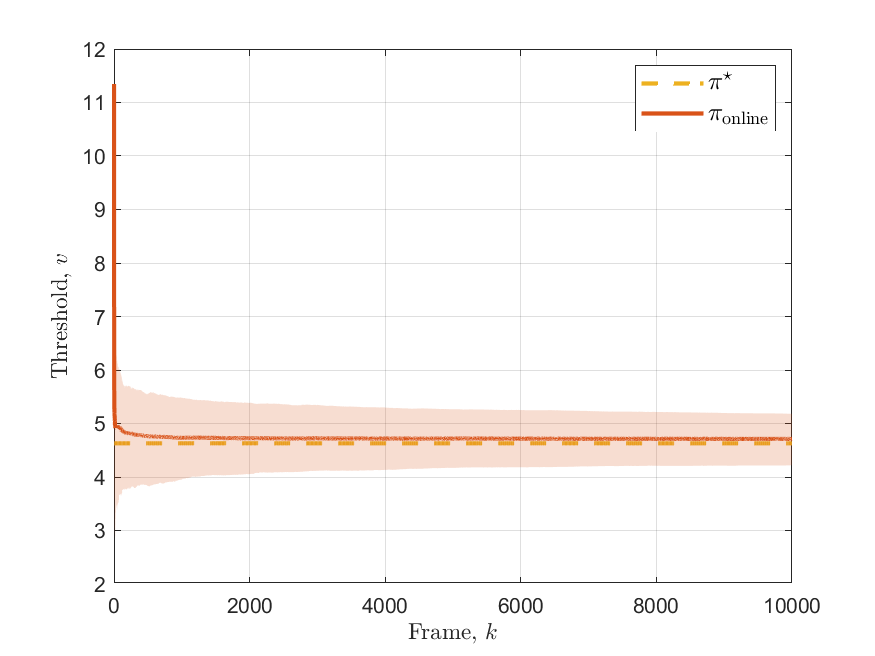}
	\caption{Threshold evolution without frequency constraint.}
	\label{Figs:thd}
\end{figure}

Next, we consider the estimation of the threshold $v(\alpha^\star - \lambda^\star)$. Obviously, the fast and accurate estimation of the threshold is the necessary condition for the convergence of MSE performance. As depicted in Fig.~\ref{Figs:thd}, the proposed algorithm can approximate the optimal threshold as the time goes to infinity. Besides, the variance of the threshold estimation will also become small, which guarantees the convergence of MSE.

\subsection{With A Sampling Frequency Constraint}
In this part, we depict the simulation results when a sampling constraint exists. The parameters of the system are the same as in Fig.~\ref{Figs:nofreq}, and we set $f_\text{max}=0.02$. In other words, the minimum average frame length $\frac{1}{f_\text{max}}=50$. Notice that now the zero-wait policy does not satisfy the sampling constraint. Therefore, we consider a frequency conservative policy $\pi_{\text{freq}}$, which selects $W_k$ as
\begin{equation*}
	W_k = \max \left\{\frac{k}{f_\text{max}} - \sum_{k'=1}^{k-1}L_{k'} - D_k,0\right\}.
\end{equation*}

\begin{figure}
	\centering
	\includegraphics[width=0.8\columnwidth]{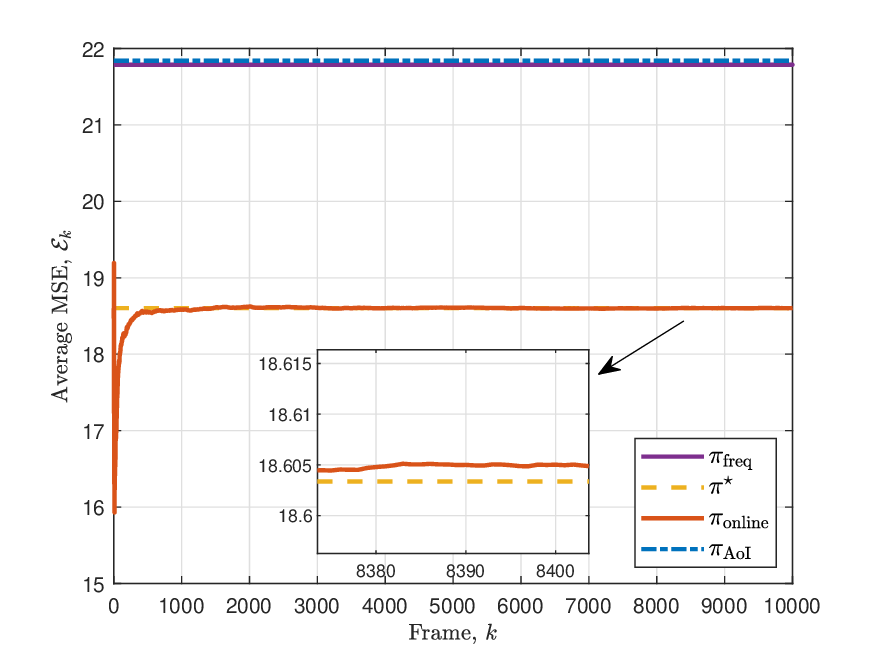}
	\caption{MSE performance under frequency constraint $f_{\text{max}}=0.02$.}
	\label{Fig:freq}
\end{figure}

\begin{figure}
	\centering
	\includegraphics[width=0.8\columnwidth]{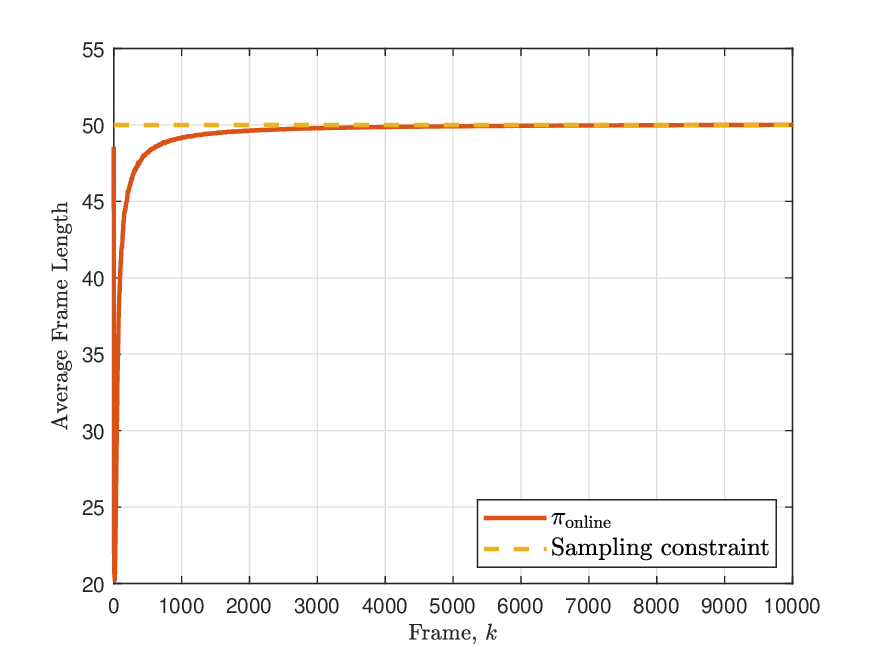}
	\caption{Average frame length under frequency constraint $f_{\text{max}}=0.02$.}
	\label{Fig:freq_con}
\end{figure}

We set the parameter $V=500$ and depict the MSE performance and average frame length in Fig.~\ref{Fig:freq} and Fig.~\ref{Fig:freq_con}. These two figures verify that the proposed algorithm can also approximate the lower bound while satisfying the frequency constraint.

\begin{figure}
	\centering
	\subfigure[MSE performance.]{
		\label{Fig:Vmse}
		\includegraphics[width=0.8\linewidth]{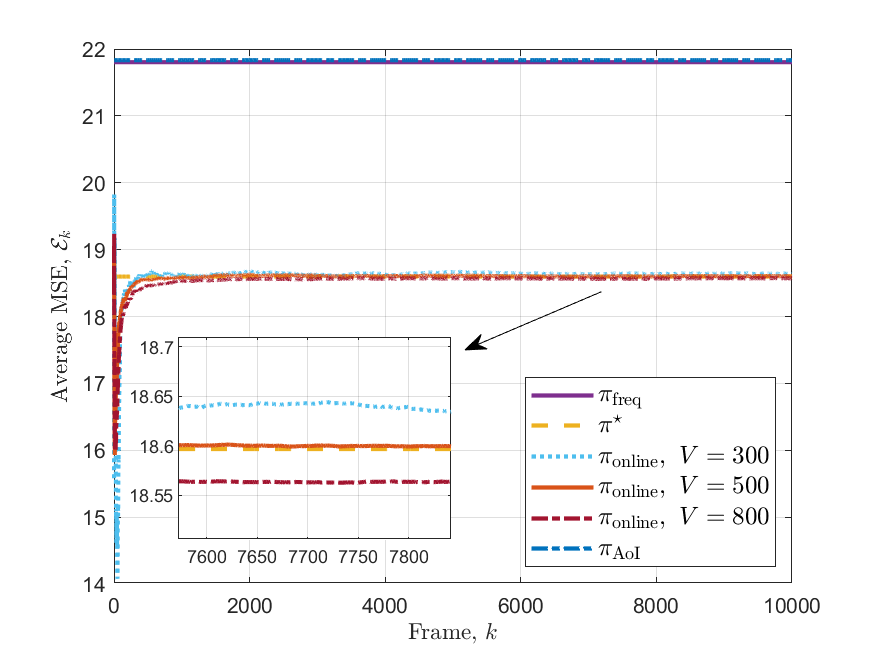}}
	\subfigure[Average frame length.]{
		\label{Fig:Vcon}
		\includegraphics[width=0.8\linewidth]{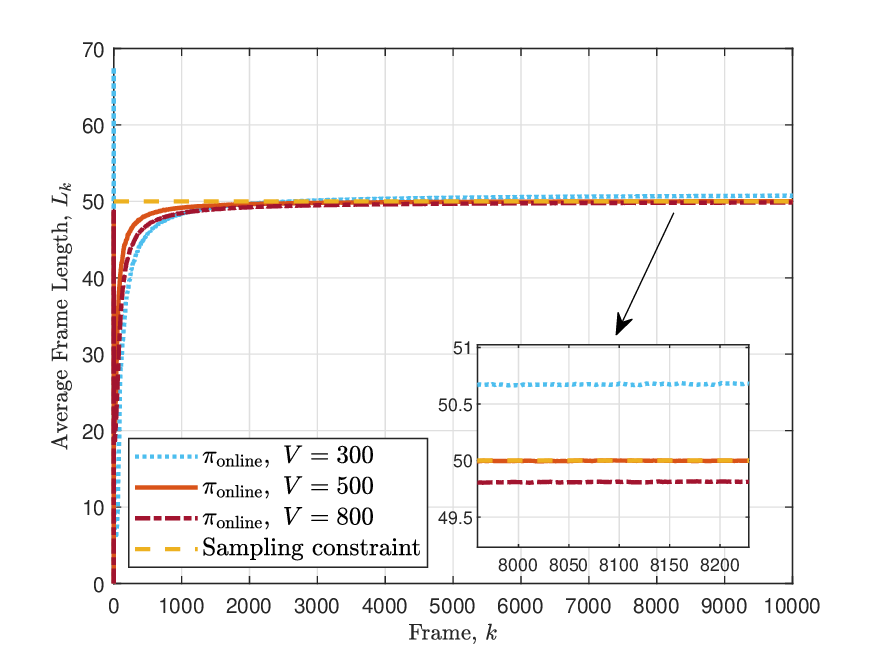}}
	\caption{MSE performance and average frame length with different parameter $V$.}
	\label{Fig:DiffV}
\end{figure}

Finally, we investigate the impact of $V$ on the MSE performance and average frame length. We choose three different values of $V=\{300,500,800\}$ and compare the MSE performance and average frame length, as depicted in Fig.~\ref{Fig:Vmse} and Fig.~\ref{Fig:Vcon} respectively. Generally speaking, the MSE performance of proposed algorithm with different $V$ can all converge to the optimal MMSE, and the average inter-update interval of the proposed algorithms are near the frequency constraint. Notice that $V$ is a hyper parameter controlling the estimation of the Lagrange multiplier. A larger $V$ indicates less emphasis on the frequency constraint. By using a larger $V=800$, the algorithm will take a longer time to converge to the sampling frequency constraint. Since for $t<8000$ the sampling frequency of the algorithm slightly violates the sampling frequency constraint, the MSE is smaller.


\section{Conclusion}
In this work, we studied the sampling policy for remote estimation of an OU process through a channel with transmission delay. We aim at designing an online sampling policy that can minimize the mean square error when the delay distribution is unknown. Finding the MSE minimum sampling policy can be reformulated into an optimal stopping problem, we proposed a stochastic approximation algorithm to learn the optimum stopping threshold adaptively. We prove that, after taking $k$ samples, the cumulative MSE regret of our proposed algorithm grows with rate $\mathcal{O}(\ln k)$, and the expected time-averaged MSE of our proposed algorithm converges to the minimum MSE almost surely. Numerical simulation validates the superiority and convergence performance of the proposed algorithm. 

\bibliographystyle{IEEEtran}
\bibliography{bibfile}

\appendices
\section{Lemmas and Notations}
First, we state the auxiliary lemmas and corollaries that will be used in the following proofs. Proofs for these lemmas and corollaries are provided in 
\begin{lemma}\label{lem:R1(v)}
	\cite[Lemma 1 Restated]{OU2021Sun}
	\begin{align}
		&\mathbb{E}[D_k + W_k] \nonumber \\
		=& \mathbb{E}[D_k] + \mathbb{E}[\max\{R_1(v((\alpha_k - \lambda_k)^+)) - R_1(|O_{D_k}|),0 \}],
	\end{align}
	where
	\begin{subequations}
		\begin{align}
			&R_1(v) = \frac{v^2}{\sigma^2} {_2F_2} \left(1,1;\frac{3}{2},2;\frac{\theta}{\sigma^2}v^2 \right), \label{eq:r1def}\\
			&{_2F_2}\left(1,1;\frac{3}{2},2;z\right) = \sum_{n=0}^\infty \frac{2^n n!n!}{(n+1)!(2n+1)!!} \frac{z^n}{n!}.
		\end{align}
	\end{subequations}
 Moreover, since $R_1(\cdot)$ is a monotonically increasing function, $v(\beta)=\frac{\sigma}{\sqrt{\theta}}G^{-1}\left(\frac{\sigma^2}{\beta}\right)$ and $G(x)=\frac{e^{x^2}}{x}\int_0^xe^{-t^2}\text{d}t$ is monotonic increasing, we have $R_1(v(\alpha))$ is monotically decreasing. 
\end{lemma}

\begin{Cor}\label{cor:lbound}
    Recall that function $l(\beta)=\mathbb{E}[D+w(O_D;\beta)]$ is the expected framelength when using sampling threshold $v(\beta)$. When there is no sampling frequency constraint and $\lambda=0$, function $l(\alpha)$ has the following property:
    \begin{equation}
        |l(\alpha)-l(\alpha^\star)|\leq N|\alpha-\alpha^\star|,\label{eq:lmapping}
    \end{equation}
    where $N=\max_{\alpha\in[\alpha_{\text{lb}}, \alpha_{\text{ub}}]}\left|R_1'(v(\alpha))v'(\alpha)\right|$ is a constant independent of $\alpha$. 
\end{Cor}
The proof is provided in Appendix~\ref{pf:lbound}
\begin{lemma}\label{lem:bound}
	Recall that $\mathbb{E}[D]\leq D_{\text{ub}}$ and $\mathbb{E}[D^2]\leq M_{\text{ub}}$ and $\alpha_k$ is truncated into interval $[\alpha_{\text{lb}}, \alpha_{\text{ub}}]$ using Lemma~\ref{lem:alpha_lb}, when there is no sampling frequency constraint and $\lambda_k\equiv 0$, we have the following bounds for each frame $k$:
	\begin{subequations}
		\begin{align}
			0 \le \mathbb{E}[O_{L_k}^2] <& \frac{\sigma^2}{2\theta}; \label{eq:Qk_ub}\\ 
			0 \le \mathbb{E}[O_{L_k}^4] <& \frac{3\sigma^4}{4\theta^2}; \label{eq:Qk2_ub}\\ 
			0 \le \mathbb{E}[L_k] \le& D_\text{ub} + \frac{v(\alpha_{\text{lb}})^2}{\sigma^2} e^{\frac{2\theta}{\sigma^2}v(\alpha_{\text{lb}})^2} \triangleq L_\text{ub};\label{eq:Lk_ub} \\ 
			0 \le \mathbb{E}[L_k^2] \le& M_\text{ub} + 2D_\text{ub}\frac{v(\alpha_{\text{lb}})^2}{\sigma^2} e^{\frac{2\theta}{\sigma^2}v(\alpha_{\text{lb}})^2} \nonumber\\
			&+\frac{2v(\alpha_{\text{lb}})^3}{\sigma^3} \sqrt{\frac{\pi}{\theta}} e^{\frac{3\theta}{\sigma^2}v(\alpha_\text{lb})^2} \triangleq L_\text{ub2}. \label{eq:Lk2_ub}
		\end{align}
	\end{subequations}
\end{lemma}

The proof of Lemma \ref{lem:bound} is provided in Appendix~\ref{pf:bound}.

\begin{lemma}\label{eq:gconcave}
For fixed $\lambda$, function $g_\lambda(\alpha)=o(\alpha-\lambda)-\alpha l(\alpha-\lambda)$ is continuous, monotonically decreasing and convex. Moreover, there exists a constant $N$ so that function $g_0(\alpha)$   
\begin{subequations}
\begin{align}
    g_0(\alpha)&\geq -l(\alpha^\star)(\alpha-\alpha^\star)+N(\alpha-\alpha^\star)^2, \\
    |g_0(\alpha)|&\leq l(\alpha^\star)|\alpha-\alpha^\star|. \label{eq:galphadiff}
\end{align}
\end{subequations}
\end{lemma}
Proof for Lemma~\ref{eq:gconcave} is provided in Appendix~\ref{pf:gconcave}. 

\begin{theorem}\label{thm:alpha}
	The estimation $\alpha_k$ computed in Algorithm \ref{alg:online} can converge to $\alpha^\star$ with probability 1, and we have
	\begin{equation}\label{eq:alpha_cov}
		\mathbb{E}[(\alpha_k-\alpha^\star)^2] \le \frac{C}{kD_\text{lb}^2}\sim \mathcal{O}\left(\frac{1}{k}\right),
	\end{equation}
	where $C$ is a constant independent of $k$, i.e., 
  \begin{align}
  C =& \frac{3\sigma^4}{4\theta^2} + \alpha_{\text{ub}}^2 \big(M_\text{ub} + 2D_\text{ub}\frac{v(\alpha_{\text{lb}})^2}{\sigma^2} e^{\frac{2\theta}{\sigma^2}v(\alpha_{\text{lb}})^2} \\
  &+ \frac{2v(\alpha_{\text{lb}})^3}{\sigma^3} \sqrt{\frac{\pi}{\theta}} e^{\frac{3\theta}{\sigma^2}v(\alpha_\text{lb})^2}\big). \label{eq:Cdef}
  \end{align}
 \end{theorem}

The proof of Theorem~\ref{thm:alpha} is the same as \cite[Lemma 6]{JMLR2021neely}. 
\section{Proof of Lemma \ref{lemma:RR}}\label{pf:lemmaRR}

	The ultimate goal is to rewrite the averaged MMSE \eqref{eq:primalobj} obtained by a stationary policy as the time-averaged cost of each frame. The waiting time $W_k$ set by any stationary policy $\pi$ can be viewed as a stopping time. The information, i.e., tuple $\{(D_k, \Delta_{S_{k+1}})\}$ is a regenerative sequence as the instant estimation error $\Delta_t, t\geq S_{k}+D_k$ is an OU process starting from time $t=S_k$. Therefore, for stationary policy, the cumulative estimation error in frame $k$, i.e., $E_k:=\int_{S_k}^{S_{k+1}}(X_t-\hat{X}_t)^2\text{d}t$ and $L_k:=S_{k+1}-S_k$ are generative random processes. Then according the renewal-reward theory \cite{ross2013applied}, both the average cumulative MSE in each frame $\{\frac{1}{K}\mathbb{E}\left[\sum_{k=1}^KE_k\right]\}$ and the average frame-length  $ \{\frac{1}{K}\mathbb{E}\left[\sum_{k=1}^KL_k\right]\}$ have limits. Then according to the renewal reward theory \cite{ross2013applied}, the time averaged MMSE can be computed by:
	\begin{align}
		&\limsup_{T\rightarrow\infty}\frac{1}{T}\mathbb{E}\left[\int_{t=0}^T\left(X_t-\hat{X}_t\right)^2\text{d}t\right]\nonumber\\
{=}&\limsup_{K\rightarrow\infty}\frac{\sum_{k=1}^K\mathbb{E}\left[\int_{S_k}^{S_{k+1}}(X_t-\hat{X}_t)^2\text{d}t\right]}{\sum_{k=1}^K\mathbb{E}\left[\left(S_{k+1}-S_k\right)\right]}.\label{eq:frame-mse}
	\end{align}
	
Then to compute the average cost in each frame $k$, we introduce the following properties of the stopping time of an OU process:
	\begin{lemma}[Lemma 5, \cite{OU2021Sun} Restated]
		\label{lemma:4-2}Let $O_t$ be an OU process with initial state zero and parameter $\mu=0$, and $\tau$ is a stopping time with $\mathbb{E}[\tau]<\infty$, the integral of $O_t^2$ from $0$ to $t$ can be computed by
		\begin{equation}
\mathbb{E}\left[\int_{0}^{\tau}O_t^2\text{d}t\right]=\mathbb{E}\left[\frac{\sigma^2}{2\theta}\tau-\frac{1}{2\theta}O_{\tau}^2\right]. 
		\end{equation}
  \label{lemma:OU-err}
	\end{lemma}
	
	We then proceed to compute the expected cumulative error of stationary policy $\pi$ using Lemma~\ref{lemma:OU-err}. Notice that the interval $[S_k, S_{k+1})$ can then be divided into two intervals $[S_k, S_k+D_k)$ and $[S_k+D_k, S_k+D_k+W_k)$. The cumulative estimation error during $[S_k, S_k+D_k)$ can be computed as follows:
	\begin{align}
		&\mathbb{E}\left[\int_{S_k}^{S_k+D_k}(X_t-\hat{X}_t)^2\text{d}t\right]\nonumber\\
		=&\mathbb{E}\left[\int_{S_{k-1}}^{S_{k-1}+D_{k-1}+W_{k-1}+D_k}(X_t-\hat{X}_t)^2\text{d}t\right]\nonumber\\
  &-\mathbb{E}\left[\int_{S_{k-1}}^{S_{k-1}+D_{k-1}+W_{k-1}}(X_t-\hat{X}_t)^2\text{d}t\right]\nonumber\\
		\overset{(a)}{=}&\mathbb{E}\left[\frac{\sigma^2}{2\theta}\left(D_{k-1}+W_{k-1}+D_k\right)-\frac{1}{2\theta}O_{D_{k-1}+W_{k-1}+D_k}^2\right]\nonumber\\
  &-\mathbb{E}\left[\frac{\sigma^2}{2\theta}\left(D_{k-1}+W_{k-1}\right)-\frac{1}{2\theta}O_{D_{k-1}+W_{k-1}}^2\right],\label{eq:rr-1}
	\end{align}
 where equation $(a)$ is because during interval $[S_k, S_k+D_k)$, the instant $X_t-\hat{X}_t$ from \eqref{eq:err} is equivalent to an OU process starting from time $t=S_{k-1}$, and the cumulative MSE can be computed by Lemma~\ref{lemma:OU-err}. Notice that the delay distribution $D_k$ is independent of $O_{D_{k-1}+W_{k-1}}$. Therefore, 
 \begin{align}
     &\mathbb{E}\left[O_{D_{k-1}+W_{k-1}+D_k}^2\right]\nonumber\\
     =&\mathbb{E}\left[\left(O_{D_{k-1}+W_{k-1}}e^{-\theta D_k}+\frac{\sigma}{\sqrt{2\theta}}e^{-\theta D_k}W_{e^{2\theta D_k}-1}\right)^2\right]\nonumber\\
     =&\mathbb{E}[O_{D_{k-1}+W_{k-1}}^2]\mathbb{E}[e^{-2\theta D_k}]+\frac{\sigma^2}{2\theta}\mathbb{E}\left[1-e^{-2\theta D_k}\right]. \label{eq:odwd}
 \end{align}
 Plugging \eqref{eq:odwd} into \eqref{eq:rr-1}, we have:
 	\begin{align}
		&\mathbb{E}\left[\int_{S_k}^{S_k+D_k}(X_t-\hat{X}_t)^2\text{d}t\right]\nonumber\\
{=}&\mathbb{E}\left[\frac{\sigma^2}{2\theta}\left(D_{k-1}+W_{k-1}+D_k\right)\right]\nonumber\\
&-\frac{1}{2\theta}\mathbb{E}[O_{D_{k-1}+W_{k-1}}^2]\mathbb{E}[e^{-2\theta D_k}]-\frac{\sigma^2}{4\theta^2}\mathbb{E}\left[1-e^{-2\theta D_k}\right]\nonumber\\
  &-\mathbb{E}\left[\frac{\sigma^2}{2\theta}\left(D_{k-1}+W_{k-1}\right)-\frac{1}{2\theta}O_{D_{k-1}+W_{k-1}}^2\right],\label{eq:rr-1prime}
	\end{align}
 Similarly, the second part of the cumulative MSE, i.e., the cumulative MSE during interval $[S_k+D_k, S_k+D_k+W_k)$ can be computed by
	\begin{align}
		&\mathbb{E}\left[\int_{S_k+D_k}^{S_k+D_k+W_k}(X_t-\hat{X}_t)^2\text{d}t\right]\nonumber\\
		=&\mathbb{E}\left[\int_{S_k}^{S_k+D_k+W_k}(X_t-\hat{X}_t)^2\text{d}t\right]\nonumber\\
  &-\mathbb{E}\left[\int_{S_k}^{S_k+D_k}(X_t-\hat{X}_t)^2\text{d}t\right]\nonumber\\
		\overset{(b)}{=}&\mathbb{E}\left[\frac{\sigma^2}{2\theta}\left(D_k+W_k\right)-\frac{1}{2\theta}O_{D_{k}+W_{k}}^2\right]\nonumber \\
  &-\mathbb{E}\left[\frac{\sigma^2}{2\theta}D_k-\frac{1}{2\theta}O_{D_{k}}^2\right],\label{eq:rr-2}
	\end{align}
 where equation $(b)$ is obtained because the instant estimation error $X_t-\hat{X}_t, t\geq S_k+D_k$ is an OU process starting at time $S_k$ according to \eqref{eq:err}. 
	
	By summing up \eqref{eq:rr-1prime} and \eqref{eq:rr-2}, we are able to compute the expected cumulative error for stationary policy $\pi$:
	\begin{align}
 &\mathbb{E}\left[E_k\right]=\mathbb{E}\left[\int_{S_k}^{S_{k+1}}(X_t-\hat{X}_t)^2\text{d}t\right]\nonumber\\
		=&\mathbb{E}\left[\frac{\sigma^2}{2\theta}\left(D_{k-1}+W_{k-1}+D_k\right)\right]\nonumber\\
&-\frac{1}{2\theta}\mathbb{E}[O_{D_{k-1}+W_{k-1}}^2]\mathbb{E}[e^{-2\theta D_k}]-\frac{\sigma^2}{4\theta^2}\mathbb{E}\left[1-e^{-2\theta D_k}\right]\nonumber\\
  &-\mathbb{E}\left[\frac{\sigma^2}{2\theta}\left(D_{k-1}+W_{k-1}\right)-\frac{1}{2\theta}O_{D_{k-1}+W_{k-1}}^2\right]\nonumber\\
  &+\mathbb{E}\left[\frac{\sigma^2}{2\theta}\left(D_k+W_k\right)-\frac{1}{2\theta}O_{D_{k}+W_{k}}^2\right]\nonumber \\
  &-\mathbb{E}\left[\frac{\sigma^2}{2\theta}D_k-\frac{1}{2\theta}O_{D_{k}}^2\right]\nonumber\\
\overset{(c)}{=}&\mathbb{E}\left[\frac{\sigma^2}{2\theta}(D_{k-1}+W_{k-1})\right]+\frac{1}{2\theta}\mathbb{E}\left[O_{D_k}^2\right]\nonumber\\
  &-\frac{1}{2\theta}\mathbb{E}[O_{D_{k-1}+W_{k-1}}^2]\mathbb{E}[e^{-2\theta D_k}]-\frac{\sigma^2}{4\theta^2}\mathbb{E}\left[1-e^{-2\theta D_k}\right]\nonumber\\
  \overset{(d)}{=}&\mathbb{E}\left[\frac{\sigma^2}{2\theta}(D_{k-1}+W_{k-1})\right]-\frac{1}{2\theta}\mathbb{E}[O_{D_{k-1}+W_{k-1}}^2]\mathbb{E}[e^{-2\theta D_k}],\label{eq:E-k-expect}
	\end{align}
	where equality $(c)$ is obtained because the transmission delay $D_k$ is i.i.d., and therefore
 \begin{align}
    &\mathbb{E}\left[\frac{\sigma^2}{2\theta}\left(D_{k-1}+W_{k-1}\right)-\frac{1}{2\theta}O_{D_{k-1}+W_{k-1}}^2\right]\nonumber\\
  =&\mathbb{E}\left[\frac{\sigma^2}{2\theta}\left(D_k+W_k\right)-\frac{1}{2\theta}O_{D_k+W_k}^2\right]. 
\end{align}
and equality $(d)$ is because:
\[\mathbb{E}[O_{D_k}^2]=\frac{\sigma^2}{2\theta}\mathbb{E}[1-e^{-2\theta D_k}]. \]

Finally, plugging \eqref{eq:E-k-expect} into \eqref{eq:frame-mse}, we have, with probability 1, the time-averaged MSE can be computed by:
\begin{align}  
&\limsup_{T\rightarrow\infty}\frac{1}{T}\mathbb{E}\left[\int_{t=0}^T(X_t-\hat{X}_t)^2\text{d}t\right]\nonumber\\
=&\limsup_{K\rightarrow\infty}\frac{\sum_{k=1}^K\left(\mathbb{E}\left[\frac{\sigma^2}{2\theta}(D_{k-1}+W_{k-1})\right]\right)}{\sum_{k=1}^K\mathbb{E}[D_k+W_k]}\nonumber\\
&- \frac{\sum_{k=1}^K \frac{1}{2\theta}\mathbb{E}[O_{D_{k-1}+W_{k-1}}^2]\mathbb{E}[e^{-2\theta D_k}]}{\sum_{k=1}^K\mathbb{E}[D_k+W_k]} \nonumber \\
=&-\frac{\mathbb{E}[e^{-2\theta D_k}]}{2\theta} \times\lim_{K\rightarrow\infty}\frac{\sum_{k=1}^K\mathbb{E}\left[O_{D_k+W_k}^2\right]}{\sum_{k=1}^K\mathbb{E}[D_k+W_k]}+\frac{\sigma^2}{2\theta}. \label{eq:lem1term1}
\end{align}

Notice that optimal value of LHS of \eqref{eq:lem1term1} is indeed $\text{mmse}$. Therefore, the problem is equivalent to
\begin{align*}
    &\text{mmse} \\ 
    =& \inf_{\pi\in \Pi_r} -\frac{\mathbb{E}[e^{-2\theta D_k}]}{2\theta} \times\lim_{K\rightarrow\infty}\frac{\sum_{k=1}^K\mathbb{E}\left[O_{D_k+W_k}^2\right]}{\sum_{k=1}^K\mathbb{E}[D_k+W_k]}+\frac{\sigma^2}{2\theta}
\end{align*}

Denote $\alpha^\star = \left(\frac{\sigma^2}{2\theta} - \text{mmse}\right)\frac{2\theta}{\mathbb{E}[e^{-2\theta D_k}]}$. Rearranging the terms yields
\begin{equation}
    \alpha^\star = \sup_{\pi\in \Pi_r} \lim_{K\rightarrow\infty}\frac{\sum_{k=1}^K\mathbb{E}\left[O_{D_k+W_k}^2\right]}{\sum_{k=1}^K\mathbb{E}[D_k+W_k]}.
\end{equation}

According to \cite{OU2021Sun}, we have $\text{mmse} \le \frac{\sigma^2}{2\theta}$. Therefore, $\alpha^\star \ge 0$.

\section{Proof of Lemma~\ref{lem:alpha_lb}}\label{pf:lemmaalpha}
Notice that
\begin{align*}
	\mathbb{E}[D_k + \hat{W}] > \frac{1}{f_\text{max}} + c > \frac{1}{f_\text{max}}.
\end{align*}

This means $\hat{W}$ is a fixed and feasible waiting solution to the problem. Then according to \eqref{eq:rr-obj}, we have
\begin{equation*}
	\alpha^\star \ge \frac{\mathbb{E}[O_{D+\hat{W}}^2]}{\mathbb{E}[D+\hat{W}]}.
\end{equation*}

First we bound $\mathbb{E}[D+\hat{W}] \le D_\text{ub} +\hat{W}$. Next we bound $\mathbb{E}[O_{D+\hat{W}}^2]$ as
\begin{align*}
	\mathbb{E}[O_{D+\hat{W}}^2] =& \frac{\sigma^2}{2\theta}\left(1- \mathbb{E}[e^{-2\theta(D+\hat{W})}] \right) \\
	\overset{(a)}{\ge}& \frac{\sigma^2}{2\theta}\left(1- e^{-2\theta\hat{W}} \right),
\end{align*}
where (a) holds since $D \ge 0$ and $e^{-x}$ is decreasing. Combining the above two terms we have
\begin{align*}
	\alpha^\star \ge \frac{\sigma^2(1- e^{-2\theta\hat{W}})}{2\theta (D_{\text{ub}}+\hat{W})} = \alpha_{\text{lb}}.
\end{align*}

For the upper bound, according to \cite{OU2021Sun}, we have
\begin{equation}\label{eq:mmse>mse_D}
	\text{mmse} \ge \text{mse}_D = \frac{\sigma^2}{2\theta} \mathbb{E}[1-e^{-2\theta D}].
\end{equation}

Plugging \eqref{eq:mmse>mse_D} into \eqref{eq:alpha_def} yields
\begin{equation*}
	\alpha^\star \le \left(\frac{\sigma^2}{2\theta} - \frac{\sigma^2}{2\theta} \mathbb{E}[1-e^{-2\theta D}] \right) \frac{2\theta}{\mathbb{E}[e^{-2\theta D}]} = \sigma^2= \alpha_{\text{ub}}.
\end{equation*}

\section{Proof of Lemma \ref{lemma:optstop}}\label{pf:lemma:optstop}

To solve the problem, From general optimal stopping theory \cite[Chapter 1]{peskir2006optimal}, we know that the following stopping time should be optimal:
\begin{equation}
    \tau_\star = \inf \{t\ge 0:|V_t|\ge v_\star \},
\end{equation}
where $v_\star$ is the optimal stopping threshold to be found.

We solve \eqref{eq:H(v)} by the free-boundary approach \cite{peskir2006optimal}. To find the $v_\star$, we solve the following free boundary problem:
\begin{subequations}
\begin{align}
&\frac{\sigma^2}{2}H''(v)-\theta vH'(v)=\beta, ~v\in(-v_\star,v_\star),\label{eq:free1}\\
    &H(\pm v_\star) = v_\star^2,\label{eq:free2}\\
        &H'(\pm v_\star) = \pm 2 v_\star.\label{eq:free3}
\end{align}    
\end{subequations}

where $H(v)$ is the value function of \eqref{eq:H(v)}.

Let $S(v)=H'(v)$, equation \eqref{eq:free1} implies:
\begin{equation}
    S'(v)-\frac{2\theta v}{\sigma^2} S(v)=\frac{2\beta}{\sigma^2}. \label{eq:sprime}
\end{equation}

Multiplying $e^{-\frac{\theta}{\sigma^2}v^2}$ on both sides of equation \eqref{eq:sprime}, we have:
\begin{equation}
    [S(v)e^{-\frac{\theta}{\sigma^2}v^2}]'=\frac{2\beta}{\sigma^2}e^{-\frac{\theta}{\sigma^2}v^2}. 
\end{equation}

Then 
\begin{equation}
    S(v)e^{-\frac{\theta}{\sigma^2}v^2}=C_1+\int_{0}^v\frac{2\beta}{\sigma^2}e^{-\frac{\theta}{\sigma^2}u^2}\text{d}u,
\end{equation}
where $C_1$ is a constant so that $S(\pm v_\star)$ satisfy \eqref{eq:free3}.  Denote 
\begin{equation}
    \text{erf}(x)=\frac{2}{\sqrt{\pi}}\int_0^xe^{-t^2}\text{d}t.
\end{equation}

Then,
\begin{align}
    S(v)e^{-\frac{\theta}{\sigma^2}v^2}&=C_1+\frac{2\beta}{\sigma^2}\sqrt{\frac{\pi}{4}}\frac{\sigma}{\sqrt{\theta}}\text{erf}\left(\frac{\sqrt{\theta}}{\sigma}v\right)\nonumber\\
    &=C_1+\frac{\beta}{\sigma}\sqrt{\frac{\pi}{\theta}}\text{erf}\left(\frac{\sqrt{\theta}}{\sigma}v\right)
\end{align}

Therefore, we have:
\begin{align}
    H'(v)&=S(v)=C_1e^{\frac{\theta}{\sigma^2}v^2}+\frac{\beta}{\sigma}\sqrt{\frac{\pi}{\theta}}e^{\frac{\theta}{\sigma^2}v^2}\text{erf}\left(\frac{\sqrt{\theta}}{\sigma}v\right)\nonumber\\
    &=C_1e^{\frac{\theta}{\sigma^2}v^2}+\frac{2\beta}{\sigma\sqrt{\theta}}F\left(\frac{\sqrt{\theta}}{\sigma}v\right),~v \in (-v_\star,v_\star),
\end{align}
where $F(x)=e^{x^2} \int_0^x e^{-t^2}\text{d}t$. Consider that $H'(v)$ is odd but $e^{\frac{\theta}{\sigma^2}v^2}$ is even, we have $C_1=0$. Therefore:
\begin{equation}
    H'(v)=\frac{2\beta}{\sigma\sqrt{\theta}}F\left(\frac{\sqrt{\theta}}{\sigma}v\right). \label{eq:Hd}
\end{equation}

Plugging \eqref{eq:Hd} into the boundary condition \eqref{eq:free3}, we have:
\begin{equation}
    \frac{2\beta}{\sigma\sqrt{\theta}} F\left(\frac{\sqrt{\theta}}{\sigma}v_\star\right) = 2 v_\star.\label{eq:vstar}
\end{equation}

Multiplying $\frac{\sqrt{\theta}}{\sigma}$ on both sides of \eqref{eq:vstar}, we have:
\begin{equation}
    \frac{\beta}{\sigma^2} F\left(\frac{\sqrt{\theta}}{\sigma}v_\star\right)=\frac{\sqrt{\theta}}{\sigma}v_\star.
\end{equation}

Finally, denote $G(x)=F(x)/x$. the optimum threshold $v_\star$ can be obtained by:
\begin{equation}
    \frac{\beta}{\sigma^2} G\left(\frac{\sqrt{\theta}}{\sigma}v_\star\right) =1,
\end{equation}

Therefore, we have
\begin{equation*}
    v_\star = \frac{\sigma}{\sqrt{\theta}} G^{-1} \left(\frac{\sigma^2}{\beta}\right).
\end{equation*}

\section{Proof of Theorem~\ref{thm:mmseas}}\label{pf:thmmmseas}
According to Lemma~\ref{lem:bound}, since $\alpha_k$ and $\mathbb{E}[L_k]$ is bounded by a function of $\alpha$, to show that the average MSE $\frac{1}{S_{k+1}}\int_0^{S_{k+1}}(X_t-\hat{X}_t)^2\text{d}t$ converges to mmse, it is then suffice to show that sequence 
\begin{equation}
    \xi_k:=\frac{1}{k}\left(\int_0^{S_{k+1}}(X_t-\hat{X}_t)^2\text{d}t-\text{mmse}\times S_{k+1}\right)\label{eq:thetedef}
\end{equation}
converges to 0 almost surely. 

Our proof is based on the perturbed ODE approach \cite[Chapter 7]{Kushner2003} for analyzing stochastic approximation. To use the ODE approach, first we need to rewrite $\xi_k$ in recursive form as follows:
\begin{align}
    \xi_{k}=&\frac{1}{k}\left(\int_{0}^{S_k}(X_t-\hat{X}_t)^2\text{d}t-\text{mmse}\times S_k\right.\nonumber\\
    &\left.+\int_{S_k}^{S_{k+1}}(X_t-\hat{X}_t)^2\text{d}t-\text{mmse}\times L_k\right)\nonumber\\
    \overset{(a)}{=}&\frac{1}{k}(k-1)\xi_{k-1}\nonumber\\
    &+\frac{1}{k}\left(\int_{S_k}^{S_{k+1}}(X_t-\hat{X}_t)^2\text{d}t-\text{mmse}\times L_k\right)\nonumber\\
    =&\xi_{k-1}+\frac{1}{k}\underbrace{\left(-\xi_{k-1}+\int_{S_k}^{S_{k+1}}(X_t-\hat{X}_t)^2\text{d}t-\text{mmse}\times L_k\right)}_{=:G_k},\label{eq:thetaevolve}
\end{align}
where equation $(a)$ is from the definition of $\xi_{k-1}$ in \eqref{eq:thetedef}. In equation \eqref{eq:thetaevolve}, $1/k$ can be viewed as a step-size of updating $\xi_k$ and $G_k$ is the updating direction. We can further decompose $G_{k}$ as follows:
\begin{align}
    &G_k\nonumber\\
    =&-\xi_{k-1}+\int_{S_k}^{S_k+D_k}(X_t-\hat{X}_t)^2\text{d}t\nonumber\\
    &+\int_{S_k+D_k}^{S_k+D_k+W_k}(X_t-\hat{X}_t)^2\text{d}t-\text{mmse}\times L_k\nonumber\\
    =&-\xi_{k-1}+\int_{S_k}^{S_k+D_k}(X_t-\hat{X}_t)^2\text{d}t\nonumber\\
    &+\int_{S_k+D_k}^{S_{k}+D_k+W_k}(X_t-\hat{X}_t)^2\text{d}t-\left(\frac{\sigma^2}{2\theta}-\frac{\mathbb{E}[e^{-2\theta D}]}{2\theta}\alpha^\star\right)\times L_k\nonumber\\
	=&-\xi_{k-1}+\underbrace{\int_{S_k}^{S_k+D_k}O_{L_{k-1}+(t-S_k)}^2\text{d}t}_{=:G_{k, 1}}\nonumber\\
	&+\underbrace{\int_{S_k+D_k}^{S_{k+1}}O_{D_k+(t-(S_k+D_k))}^2\text{d}t}_{=:G_{k, 2}}-\underbrace{\left(\frac{\sigma^2}{2\theta}-\frac{\mathbb{E}[e^{-2\theta D}]}{2\theta}\alpha^\star\right)\times L_k}_{=:G_{k, 3}}.\label{eq:theta-recurse}
\end{align}

Let $\mathbb{E}_k[\cdot]\triangleq\mathbb{E}[\cdot|\mathcal{H}_{k-1}]$ be the conditional probability given historical information $\mathcal{H}_{k-1}$. Then according to equation~\eqref{eq:rr-1prime}, since the transmission delay $D_k$ is independent of $O_{L_{k-1}}=X_{S_k}-\hat{X}_{S_k}$, the conditional expectation $\mathbb{E}[G_{k, 1}]$ can be computed by:
\begin{align}
		&\mathbb{E}_k[G_{k, 1}]=\mathbb{E}\left[\int_{S_k}^{S_k+D_k}(X_t-\hat{X}_t)^2\text{d}t|\mathcal{H}_{k-1}\right]\nonumber\\
{=}&\mathbb{E}\left[\frac{\sigma^2}{2\theta}D_k\right]+\frac{1}{2\theta}O_{L_{k-1}}^2\left(1-\mathbb{E}[e^{-2\theta D}]\right)\nonumber\\
&-\frac{\sigma^2}{4\theta^2}\mathbb{E}\left[1-e^{-2\theta D}\right],\label{eq:condg1}
	\end{align}

Similarly, through equation~\eqref{eq:rr-2}, the conditional expectation of the $G_{k, 2}$ can be computed by:
	\begin{align}
		&\mathbb{E}_k[G_{k, 2}]=\mathbb{E}\left[\int_{S_k+D_k}^{S_k+D_k+W_k}(X_t-\hat{X}_t)^2\text{d}t|\mathcal{H}_{k-1}\right]\nonumber\\
{=}&\mathbb{E}_k\left[\frac{\sigma^2}{2\theta}W_k-\frac{1}{2\theta}O_{L_k}^2\right]+\mathbb{E}_k\left[\frac{1}{2\theta}O_{D_{k}}^2\right]. \label{eq:condg2}
	\end{align}

 And the conditional expectation of $G_{k, 3}$ can be computed by:
 \begin{align}
     \mathbb{E}_k[G_{k, 3}]=\left(\frac{\sigma^2}{2\theta}-\frac{\mathbb{E}[e^{-2\theta D}]}{2\theta}\alpha^\star\right)\mathbb{E}_k[L_k]. \label{eq:condg3}
 \end{align}

From equations~\eqref{eq:condg1}-\eqref{eq:condg3}, we can compute the conditional expectation of $\mathbb{E}_k[G_k]$ by:
 \begin{align}
     &\mathbb{E}_k[G_k]\nonumber\\
     \overset{(b)}{=}&-\xi_{k-1}+\cancel{\mathbb{E}\left[\frac{\sigma^2}{2\theta}D_k\right]}+\frac{1}{2\theta}O_{L_{k-1}}^2\left(1-\mathbb{E}[e^{-2\theta D}]\right)\nonumber\\
     &-\cancel{\frac{\sigma^2}{4\theta^2}\mathbb{E}\left[1-e^{-2\theta D}\right]}\nonumber\\
     &+\mathbb{E}_k\left[\cancel{\frac{\sigma^2}{2\theta}W_k}-\frac{1}{2\theta}O_{L_k}^2\right]+\cancel{\mathbb{E}_k\left[\frac{1}{2\theta}O_{D_{k}}^2\right]}\nonumber\\
     &-\left(\cancel{\frac{\sigma^2}{2\theta}}-\frac{\mathbb{E}[e^{-2\theta D}]}{2\theta}\alpha^\star\right)\mathbb{E}_k[L_k]\nonumber\\
     =&-\xi_{k-1}+\frac{1-\mathbb{E}[e^{-2\theta D}]}{2\theta}O_{L_{k-1}}^2-\frac{1-\mathbb{E}[e^{-2\theta D}]}{2\theta}\alpha^\star l(\alpha_{k-1})\nonumber\\
     &-\frac{1}{2\theta}(\mathbb{E}_k[O_{L_k}^2]-\alpha_k\mathbb{E}_k[L_k])\nonumber\\
     &+\frac{1-\mathbb{E}[e^{-2\theta D}]}{2\theta}\alpha^\star l(\alpha_{k-1})\nonumber\\
     &-\frac{1}{2\theta}\alpha_k\mathbb{E}_k[L_k]+\frac{\mathbb{E}[e^{-2\theta D}]}{2\theta}\alpha^\star\mathbb{E}_k[L_k]\nonumber\\
     =&-\xi_{k-1}+\frac{1-\mathbb{E}[e^{-2\theta D}]}{2\theta}\left(O_{L_{k-1}}^2-\alpha^\star l(\alpha_{k-1})\right)\nonumber\\
     &-\frac{1}{2\theta}\left(o(\alpha_{k})-\alpha_kl(\alpha_k)\right)+\frac{1}{2\theta}\left(\alpha^\star l(\alpha_{k-1})-\alpha_kl(\alpha_k)\right)\nonumber\\
     &+\frac{\mathbb{E}[e^{-2\theta D}]}{2\theta}\alpha^\star(l(\alpha_{k})-l(\alpha_{k-1}))\nonumber\\
    =&-\xi_{k-1}-\frac{1}{2\theta}\left(o(\alpha_{k})-\alpha_kl(\alpha_k)\right)\nonumber\\
    &+\underbrace{\frac{1-\mathbb{E}[e^{-2\theta D}]}{2\theta}\left(O_{L_{k-1}}^2-o(\alpha_{k-1})\right)}_{=:\beta_{k, 1}}\nonumber\\
     &+\underbrace{\frac{1-\mathbb{E}[e^{-2\theta D}]}{2\theta}\left(o(\alpha_{k-1})-\alpha^\star l(\alpha_{k-1})\right)}_{=:\beta_{k, 2}}\nonumber\\
     &+\underbrace{\frac{1}{2\theta}\alpha^\star (l(\alpha_{k-1})-l(\alpha_k))}_{=:\beta_{k, 3}}+\underbrace{\frac{1}{2\theta}(\alpha^\star-\alpha_k)l(\alpha_k)}_{\beta_{k, 4}}\nonumber\\
     &+\underbrace{\frac{\mathbb{E}[e^{-2\theta D}]}{2\theta}\alpha^\star(l(\alpha_{k})-l(\alpha_{k-1}))}_{\beta_{k, 5}}. \label{eq:Gdecouple}
 \end{align}
 where equation $(b)$ is obtained because $\mathbb{E}\left[\frac{1}{2\theta} O_{D_k}^2\right]=\frac{\sigma^2}{4\theta^2}\mathbb{E}[1-e^{-2\theta D}]$ by equation \eqref{eq:mse_D}. Terms $\beta_{k, 1},\cdots, \beta_{k, 5}$ can be viewed as the bias terms in the ODE. Denote $\delta M_k:=G_k-\mathbb{E}_k[G_k]$ be the difference between the actual update and the conditional expectation, and define function:
 \begin{equation}
     f(\xi, \alpha)=-\xi-\frac{1}{2\theta}\left(o(\alpha)-\alpha l(\alpha)\right). 
 \end{equation}Plugging \eqref{eq:Gdecouple} into \eqref{eq:thetaevolve}, we have:
 \begin{equation}
     \xi_{k}=\xi_{k-1}+\frac{1}{k}\left(f(\xi_{k-1}, \alpha_k)+\sum_{j=1}^5\beta_{k, j}+\delta M_k\right). \label{eq:xiformal}
 \end{equation}

Denote $t_0=0$ and $t_k:=\sum_{j=0}^{k-1}\frac{1}{j}$ to be the cumulative step-size sequences. Select $m(t)\in\mathbb{N}^+$ to be the largest integer so that $t_{m(t)}\leq t$.  To show that the ODE \eqref{eq:xiformal} converges to 0 with almost surely, we will then verify the following statements, whose proof are provided in Appendix~\ref{pf:gm}:
\begin{lemma}\label{lemma:gm}
    The updating steps $\{G_k\}$ and the difference sequence $\{\delta M_k\}$ have the following properties:
    
    \textbf{(a)} For each constant $N$, the expectation $\mathbb{E}[|G_k|\mathbb{I}_{(|\xi_{k-1}|\leq N)}]$ is bounded for each $k$, i.e., 
    \begin{equation}\sup_k\mathbb{E}[|G_k|\mathbb{I}_{(|\xi_{k-1}|\leq N)}]<\infty. 
    \end{equation}

    \textbf{(b)} Function $f(\xi, \alpha)$ is continuous in $\xi$ for each $\alpha$. 

    \textbf{(c)} For any running time $T$, the following limit holds for all $\xi$ and $\mu>0$:
    \begin{align}
        \lim_{k\rightarrow\infty}\text{Pr}\left(\sup_{j\geq k}\max_{0\leq t\leq T}\left|\sum_{i=m(jT)}^{m(jT+t)-1}\frac{1}{i}(f(\xi, \alpha_i)-f(\xi, \alpha^\star)\right|\geq \mu\right)\nonumber\\
        =0. \nonumber
    \end{align}

    \textbf{(d)} The difference sequence $\delta M_k=G_k-\mathbb{E}_k[G_k]$ satisfies:
    \begin{equation}
        \lim_{k\rightarrow\infty}\text{Pr}\left(\sup_{j\geq k}\max_{0\leq t\leq T}\left|\sum_{i=k}^j\frac{1}{i}\delta M_i\right|\geq \mu\right)=0. \nonumber. 
    \end{equation}

    \textbf{(e)} The sum of the bias terms defined in \eqref{eq:Gdecouple} satisfies:
    \begin{equation}
        \lim_{k\rightarrow\infty}\text{Pr}\left(\sup_{j\geq k}\max_{0\leq t\leq T}\left|\sum_{i=m(jT)}^{m(jT+t)-1}\sum_{b=1}^5\frac{1}{i}\beta_{i, b}\right|\geq \mu\right)=0. 
    \end{equation}

    \textbf{(f)} Function $f(\xi, \alpha)$ can be decomposed into the sum of function of $\xi$ and a function of $\alpha$, i.e., 
    \begin{equation}
        f(\xi, \alpha)=-\xi-\frac{1}{2\theta }g_0(\alpha). 
    \end{equation}
    Since $g_0(\alpha^\star)=0$, we have $-\xi=f(\xi, \alpha^\star)$. Moreover, 
    \begin{equation}
        \lim_{k\rightarrow\infty}\text{Pr}\left(\sup_{j\geq n}\sum_{i=m(j\tau)}^{m(j\tau+\tau)-1}|\frac{1}{i}g_0(\alpha_i)|\geq \mu\right)=0. 
    \end{equation}

    \textbf{(g)} For each $\xi, \xi'$, function $f(\xi, \alpha)$ satisfies:
    \begin{equation}
        |f(\xi, \alpha)-f(\xi', \alpha)|=|\xi-\xi'|. 
    \end{equation}
\end{lemma}
Finally, according to \cite[p. 166, Theorem 1.1]{Kushner2003}, sequence $\{\xi_k\}$ converges to some limits of the ODE:
\begin{equation}
    \dot{\xi}=f(\xi, \alpha^\star)=-\xi. \label{eq:ode}
\end{equation}
Since function $f(\cdot, \alpha^\star)$ is monotonically decreasing, $\xi=0$ is the unique equilibrium point of the ODE~\eqref{eq:ode}. Therefore, $\xi_k$ converges to 0 almost surely, and the time-averaged MSE converges to the mmse with probability 1. 
\section{Proof of Theorem~\ref{thm:mmse}}\label{pf:thmmmse}
The cumulative regret, i.e., the difference between the expected cumulative MSE using the online algorithm compared with the MSE optimum sampling up to sample $(K+1)$ can be upper bounded as follows:
\begin{align}
    \mathcal{R}_K=&\mathbb{E}\left[\int_0^{S_{K+1}}(X_t-\hat{X}_t)^2\text{d}t\right]-\text{mmse}\times\mathbb{E}[S_{k+1}]\nonumber\\
    \overset{(a)}{=}&-\frac{\mathbb{E}[e^{-2\theta D_k}]}{2\theta}\times\left(\sum_{k=1}^K\mathbb{E}[O_{D_k+W_k}^2]\right)\nonumber\\
    &+(\text{mse}_{\infty}-\text{mmse})\times\left(\sum_{k=1}^K\mathbb{E}[L_k]\right)\nonumber\\
    \overset{(b)}{=}&\frac{\mathbb{E}[e^{-2\theta D_k}]}{2\theta}\times\left(-\sum_{k=1}^K\left(\mathbb{E}[O_{D_k+W_k}^2]-\alpha^\star\mathbb{E}[L_k]\right)\right)\label{eq:reg1},
\end{align}
where equation $(a)$ is obtained by \eqref{eq:lem1term1} and $\text{mse}_{\infty}=\frac{\sigma^2}{2\theta}$, and equation $(b)$ is obtained by substituting $\text{mse}_{\infty}-\text{mmse}=\alpha^\star\frac{\mathbb{E}[e^{-2\theta D}]}{2\theta}$ from equation \eqref{eq:alpha_def}. 

Then to further bound the cumulative regret computed by \eqref{eq:reg1}, let $W_k^\star$ be the waiting time selected by using parameter $\alpha^\star$ (i.e., the MSE minimum sampling policy). Then it is suffice to upper bound each term $-\mathbb{E}[O_{D_k+W_k}^2]+\alpha^\star\mathbb{E}[L_k]$ for each $k$ as follows: 
\begin{align}
    &-\mathbb{E}[O_{D_k+W_k}^2-\alpha^\star L_k]\nonumber\\
    =&-\mathbb{E}[O_{D_k+W_k}^2-\alpha_kL_k]-\mathbb{E}[(\alpha_k-\alpha^\star)L_k]\nonumber\\
    \overset{(c)}{\leq}&-\mathbb{E}[O_{D_k+W_k^\star}^2-\alpha_kL_k^\star]-\mathbb{E}[(\alpha_k-\alpha^\star)L_k]\nonumber\\
    =&-\mathbb{E}[O_{D_k+W_k^\star}^2-\alpha^\star L_k^\star]-\mathbb{E}[(\alpha_k-\alpha^\star)(l(\alpha_k)-l(\alpha^\star))]\nonumber\\
    \overset{(d)}{=}&-\mathbb{E}[(\alpha_k-\alpha^\star)(l(\alpha_k)-l(\alpha^\star)]\nonumber\\
    \overset{(e)}{\leq}&\max_{\alpha\in[\alpha_{\text{lb}}, \alpha_{\text{ub}}]}\left|R_1'(v(\alpha))v'(\alpha)\right|\times\left|\alpha_k-\alpha^\star\right|. \label{eq:regeachub}
\end{align}
where equation $(c)$ is because $W_k$ is the optimum policy that minimizes $-\mathbb{E}[O_{D_k+w}^2]+\alpha_k\mathbb{E}[D_k+w]$ and therefore we have $-\mathbb{E}[O_{D_k+W_k}^2-\alpha_kL_k]\leq-\mathbb{E}[O_{D_k+W_k^\star}^2-\alpha^\star L_k^\star]$; equation $(d)$ is because $\mathbb{E}[O_{D_k+W_k^\star}^2-\alpha^\star L_k^\star]=0$ by equation \eqref{eq:opt_b}; equation $(e)$ is from Corollary~\ref{cor:lbound}. 

Finally, plugging inequality \eqref{eq:regeachub} into \eqref{eq:reg1} for each term $k$, the cumulative regret $\mathcal{R}_K$ can be bounded, i.e.,
\begin{align}
    \mathcal{R}_K \overset{(f)}{\leq}&\frac{\mathbb{E}[e^{-2\theta D}]}{2\theta}\times\left(\sum_{k=1}^K\frac{C}{D_{\text{lb}}^2}\max_{\alpha\in[\alpha_{\text{lb}}, \alpha_{\text{ub}}]}|R_1'(v(\alpha))v'(\alpha)|\frac{1}{k}\right)\nonumber\\
    \leq& \frac{\mathbb{E}[e^{-2\theta D}]}{2\theta}\frac{C}{D_{\text{lb}}^2}\max_{\alpha\in[\alpha_{\text{lb}}, \alpha_{\text{ub}}]}|R_1'(v(\alpha))v'(\alpha)|\ln (K+1),
\end{align}
where equation $(f)$ is obtained by Theorem \ref{thm:alpha}.

\section{Proof of Lemma~\ref{lemma:gm}}\label{pf:gm}
We will verify each statement in Lemma~\ref{lemma:gm} respectively:

\textbf{(a)} By substituting $G_k$ with equation~\eqref{eq:theta-recurse}, we can upper bound $\mathbb{E}[|G_k|\mathbb{I}_{(|\xi_{k-1}|\leq N)}]$ as follows:
\begin{align}
    &\mathbb{E}[|G_k|\mathbb{I}_{(|\xi_{k-1}|\leq N)}]\nonumber\\
    \leq&\mathbb{E}[|\xi_{k-1}|\mathbb{I}_{(|\xi_{k-1}|\leq N}]+\mathbb{E}\left[\int_{t=S_k}^{S_{k+1}}(X_t-\hat{X}_t)^2\text{d}t\right]\nonumber \\
    &+\text{mmse}\mathbb{E}[L_k]. 
\end{align}
The first term $\mathbb{E}[|\xi_{k-1}|\mathbb{I}_{(|\xi_{k-1}|\leq N}]\leq N<\infty$ is bounded. Then notice that $\mathbb{E}[L_k]$ is bounded by Lemma \ref{lem:bound} and $\text{mmse}\le \text{mse}_\infty$, the third term $\text{mmse}\mathbb{E}[L_k]$ is also bounded. It then remains to show that the second term $\mathbb{E}\left[\int_{S_k}^{S_{k+1}}(X_t-\hat{X}_t)^2\text{d}t\right]$ is bounded. According to \eqref{eq:E-k-expect}, the expectation of the second term can be computed by:
\begin{align}
    &\mathbb{E}\left[\int_{S_k}^{S_{k+1}}(X_t-\hat{X}_t)^2\text{d}t\right]\nonumber\\
    =&\mathbb{E}\left[\frac{\sigma^2}{2\theta}L_{k-1}\right]-\frac{1}{2\theta}\mathbb{E}[O_{L_{k-1}}^2]\mathbb{E}[e^{-2\theta D_k}]. 
\end{align}
Since $\alpha_k\in[\alpha_\text{lb}, \alpha_{\text{ub}}]$ is bounded and function $l(\alpha_{k-1})=\mathbb{E}[L_{k-1}]$, $o(\alpha_{k-1})=\mathbb{E}[O_{L_{k-1}}^2]$ are both bounded for $\alpha\in[\alpha_\text{lb}, \alpha_{\text{ub}}]$, the expectation of the second term $\mathbb{E}[\int_{S_k}^{S_{k+1}}(X_t-\hat{X}_t)^2\text{d}t]$ is also bounded. This verifies statement $(a)$. 

\textbf{(b)} Function $f(\xi, \alpha)$ can be decoupled into $=-\xi - \frac{1}{2\theta}g_0(\alpha)$ and is thus continuous in $\xi$ for each $\alpha$. 

To proceed with the proof of statement $(c)-(f)$, we re-state the following lemma, whose proof is provided in \cite[Appendix G]{thy2022wiener}
\begin{lemma}\label{lemma:psiconverge}
    Let $\{\psi_k\}$ be a sequence. Then $\lim_{k\rightarrow\infty}\text{Pr}\left(\sup_{j\geq k}\left|\sum_{i=k}^j\frac{1}{i}\psi_i\right|\geq\mu\right)=0$ holds if one of the following condition is satisfied:
    \begin{itemize}
        \item[(1)] $\psi_k$ is a martingale sequence and its second order moment is bounded, i.e., $\mathbb{E}_k[\psi_k]=0, \sup_k\mathbb{E}[\psi_k^2]<\infty$. The correlation between each $(k, k'), k\neq k'$ pair satisfies: $\mathbb{E}[\psi_k\psi_{k'}]=0$. 
        \item[(2)] $\mathbb{E}[|\psi_k|]=\mathcal{O}(k^{-\varepsilon}), \varepsilon>0$. 
    \end{itemize}
\end{lemma}
\textbf{(c)} According to Lemma~\ref{eq:gconcave}, since $g_0(\alpha)$ is monotonic decreasing and convex, the difference $|g_0(\alpha)-g_0(\alpha')|\leq N_1|\alpha-\alpha'|$. Therefore, 
\begin{align}
    |f(\xi, \alpha_k)-f(\xi, \alpha^\star)|&=\frac{1}{2\theta}|g_0(\alpha_k)-g_0(\alpha^\star)|\nonumber \\
    &\leq \frac{N_1}{2\theta}|\alpha_k-\alpha^\star|.
\end{align}

Therefore, the expectation of $f(\xi, \alpha_k)-f(\xi, \alpha^\star)$ can be upper bounded by:
\begin{align}
    &\mathbb{E}\left[f(\xi, \alpha_k)-f(\xi, \alpha^\star)\right]\leq\mathbb{E}[\frac{N_1}{2\theta}|\alpha_k-\alpha^\star|]\nonumber\\
    \overset{(a)}{\leq}&\frac{N_1}{2\theta}\sqrt{\mathbb{E}[(\alpha_k-\alpha^\star)^2]}\overset{(b)}{=}\frac{N_1}{2\theta}\sqrt{\frac{C}{D_{\text{lb}}^2}}\sqrt{\frac{1}{k}}. 
\end{align}
where equality $(a)$ is by Cauchy-Schwartz inequality and equality $(b)$ is from Theorem~\ref{thm:alpha}. Since term $f(\xi, \alpha_k)-f(\xi, \alpha^\star)$ satisfies condition 2 in Lemma~\ref{lemma:psiconverge}, statement (c) is verified. 

\textbf{(d)} Denote $\delta M_{k, j}:=G_{k, j}-\mathbb{E}_k[G_{k, j}]$. Since $G_k=G_{k, 1}+G_{k, 2}-G_{k, 3}$, the difference term $\delta M_k=\delta M_{k, 1}+\delta M_{k, 2}-\delta M_{k, 3}$ also consists of three parts. By the union bound, 
\begin{align}
    &\lim_{k\rightarrow\infty}\text{Pr}\left(\sup_{j\geq k}\max_{0\leq t\leq T}\left|\sum_{i=k}^j\frac{1}{i}\delta M_i\right|\geq\mu\right)\nonumber\\
    \leq&\sum_{p=1}^3\lim_{k\rightarrow\infty}\text{Pr}\left(\sup_{j\geq k}\max_{0\leq t\leq T}\left|\sum_{i=k}^j\frac{1}{i}\delta M_{i, p}\right|\geq\mu/3\right).\label{eq:deltaMdecompose}
\end{align}

Therefore, to show that statement (d) is satisfied, it is suffice to show that each term $\delta M_{i, p}, p=1, 2, 3$ satisfies condition (1) in Lemma~\ref{lemma:psiconverge}. 

Notice that for fixed $O_{L_{k-1}}$, the first difference term $\delta M_{k, 1}=G_{k, 1}-\mathbb{E}_k[G_{k, 1}]$ depends only on $D_k$ and the OU process evolution during $[S_k, S_k+D_k)$. Therefore, $\mathbb{E}[\delta M_{k, 1}]=0$ and $\mathbb{E}[\delta M_{k, 1}\delta M_{k', 1}]=0, \forall k\neq k'$ due to the independence of $D_k$ and $D_{k'}$. Then, notice that $\text{Var}(\delta M_{k, 1})\leq \mathbb{E}[\delta M_{k, 1}^2]\leq \mathbb{E}[G_{k, 1}^2]$. To show that $\text{Var}(\delta M_{k, 1})<\infty$ is bounded, it is suffice to show $\mathbb{E}[G_{k, 1}^2]$ is bounded, which is shown as follows:
\begin{align}
    &\mathbb{E}[G_{k, 1}^2]\nonumber\\
    =&\mathbb{E}\left[\left(\int_{t=S_k}^{S_k+D_k}O_{L_{k-1}+(t-S_k)}^2\text{d}t\right)^2\right]\nonumber\\
    \overset{(b)}{\leq}&\mathbb{E}\left[D_k\int_{t=S_k}^{S_k+D_k}O_{L_{k-1}+(t-S_k)}^4\text{d}t\right]\nonumber\\
    \overset{(c)}{\leq}&\mathbb{E}\left[D_k\int_{t=S_k}^{S_k+D_k}3\left(\frac{\sigma^2}{2\theta}(1-e^{-2\theta(L_{k-1}+(t-S_{k}))})\right)^2\text{d}t\right]\nonumber\\
    \leq&3D_{\textbf{ub}}\left(\frac{\sigma^2}{2\theta}\right)^2. 
\end{align}
where equation $(b)$ is by Cauchy-Schwartz inequality; inequality $(c)$ is from \eqref{eq:OL4ub}. Since $\delta M_{k, 1}$ meets the first condition in Lemma~\ref{lemma:psiconverge}, we have:
\begin{equation}
    \lim_{k\rightarrow\infty}\text{Pr}\left(\sup_{j\geq k}\max_{0\leq t\leq T}\left|\sum_{i=k}^j\frac{1}{i}\delta M_{i, 1}\geq \frac{1}{3}\mu\right|\right)=0. 
\end{equation}

The difference sequence $\delta M_{k, 2}$ and $\delta M_{k, 3}$ only depends on the transmission delay $D_k$ and the OU process evolution in frame $k$. Using similar methods, it can be shown that sequences $\{\delta M_{k, 2}\}$ and $\{\delta M_{k, 3}\}$ satisfy condition 1 in Lemma~\ref{lemma:psiconverge}. Since $\lim_{k\rightarrow\infty}\text{Pr}\left(\sup_{j\geq k}\max_{0\leq t\leq T}\left|\sum_{i=k}^j\frac{1}{i}\delta M_{i, p}\geq \frac{1}{3}\mu\right|\right)=0$ holds for $p=1, 2, 3$, plugging into inequality \eqref{eq:deltaMdecompose} verifies statement (d). 

\textbf{(e)} Through the union bound, we have:
\begin{align}
    &\text{Pr}\left(\sup_{j\geq k}\max_{0\leq t\leq T}\left|\sum_{i=k}^j\frac{1}{i}\sum_{b=1}^5\beta_{i, b}\right|\geq \mu\right)\nonumber\\
    \leq&\sum_{b=1}^5\text{Pr}\left(\sup_{j\geq k}\max_{0\leq t\leq T}\left|\sum_{i=k}^j\frac{1}{i}\beta_{i, p}\right|\geq \mu/5\right).\label{eq:betaunion} 
\end{align}

To show that statement (e) holds, it is suffice to show that each of the bias term satisfy:
\begin{align}
    \lim_{k\rightarrow\infty}\text{Pr}\left(\sup_{j\geq k}\max_{0\leq t\leq T}\left|\sum_{i=m(jT)}^{m(jT+t)-1}\frac{1}{i}\delta \beta_{i, p}\right|\geq\mu/5\right)=0,\nonumber\\
    \forall p.\label{eq:betaasp}
\end{align}the second condition in Lemma~\ref{lemma:psiconverge}. We will then upper bound the expectation of each bias term $\mathbb{E}[\beta_{k, p}]$ respectively. 

The first bias term satisfies $\mathbb{E}[\beta_{k, 1}]=0$ and is hence a martingale sequence. We can bound $\mathbb{E}[\beta_{k,1}^2]$ by:
\begin{align}
    &\mathbb{E}[\beta_{k, 1}^2]=\text{Var}[\beta_{k, 1}]\nonumber\\
    =&\text{Var}\left[\left(\frac{1-\mathbb{E}[e^{-2\theta D}]}{2\theta}(O_{L_{k-1}}^2-o(\alpha_{k-1}))\right)\right]\nonumber\\
    =&\mathbb{E}\left[\left(\frac{1-\mathbb{E}[e^{-2\theta D}]}{2\theta}(O_{L_{k-1}}^2-o(\alpha_{k-1}))\right)^2\right]\nonumber\\
    \leq&2\left(\frac{1-\mathbb{E}[e^{-2\theta D}]}{2\theta}\right)^2\mathbb{E}\left[O_{L_{k-1}}^4+o(\alpha_{k-1})^2\right].
\end{align}

Then according to Lemma~\ref{lem:bound}, $\mathbb{E}[L_{k-1}^4]<\infty$ and $\mathbb{E}[o(\alpha_{k-1})^2]=\mathbb{E}[O_{L_{k-1}}^2]^2<\infty$, term $\beta_{k, 1}$ satisfies Condition~1, Lemma~\ref{lemma:psiconverge}. Therefore, equation~\eqref{eq:betaasp} holds for $p=1$
. 
The expectation of the second bias term $\beta_{k, 2}$ can be upper bounded by:
\begin{align}
    &\mathbb{E}\left[|\beta_{k, 2}|\right]\nonumber\\
    =&\frac{1-\mathbb{E}[e^{-2\theta D}]}{2\theta}\mathbb{E}[|o(\alpha_{k-1})-\alpha^\star l(\alpha_{k-1})|]\nonumber\\
    =&\frac{1-\mathbb{E}[e^{-2\theta D}]}{2\theta}\left(\mathbb{E}[|o(\alpha_{k-1})-\alpha_{k-1}l(\alpha_{k-1})|]\right.\nonumber\\
    &\left.+\mathbb{E}[(\alpha_{k-1}-\alpha^\star)l(\alpha_{k-1})]\right)\nonumber\\
    \leq&\frac{1-\mathbb{E}[e^{-2\theta D}]}{2\theta}\mathbb{E}[N_1\times |\alpha_{k-1}-\alpha^\star|]\nonumber\\
    &+\frac{1-\mathbb{E}[e^{-2\theta D}]}{2\theta}l(\alpha_{\text{lb}})\mathbb{E}[|\alpha_{k-1}-\alpha^\star|].\label{eq:betaub}
\end{align}

Recall that by Theorem~\ref{thm:alpha}, $\mathbb{E}[|\alpha_{k-1}-\alpha^\star|]\leq\sqrt{\mathbb{E}[(\alpha_{k-1}-\alpha^\star)^2]}=\mathcal{O}(1/\sqrt{k})$, equation~\eqref{eq:betaub} implies $\mathbb{E}[|\beta_{k, 1}|]=\mathcal{O}(1/\sqrt{k})$ and satisfies Lemma~\ref{lemma:psiconverge} condition 2. Equation~\eqref{eq:betaasp} holds for $p=2$. 

We then proceed to upper bound the expectation of of the third bias term by:
\begin{align}
    &\mathbb{E}[|\beta_{k, 2}|]=\mathbb{E}\left[\frac{1}{2\theta}\alpha^\star(l(\alpha_{k-1})-l(\alpha_k))\right]\nonumber\\
    \leq&\frac{1}{2\theta}\alpha^\star\left(\mathbb{E}\left[|l(\alpha_{k-1})-l(\alpha^\star)|\right]+\mathbb{E}\left[|l(\alpha_k)-l(\alpha^\star)|\right]\right)\nonumber\\
    \overset{(d)}{\leq}&\frac{1}{2\theta}\alpha^\star N\left(\mathbb{E}[|\alpha_{k-1}-\alpha^\star|]+\mathbb{E}[|\alpha_k-\alpha^\star|]\right)\nonumber\\
    \leq&\frac{1}{2\theta}\alpha^\star N\left(\sqrt{\mathbb{E}[|\alpha_{k-1}-\alpha^\star|^2]}+\sqrt{\mathbb{E}[|\alpha_k-\alpha^\star|^2]}\right)\nonumber\\
    =&\mathcal{O}(k^{-1/2}). 
\end{align}
where inequality $(d)$ is obtained by Corollary~\ref{cor:lbound}. Therefore, $\beta_{k, 2}$ also satisfies the Condition 2 in Lemma~\ref{lemma:psiconverge} and equation~\eqref{eq:betaasp} holds for $p=2$. Since term $\alpha_k\in[\alpha_{\text{lb}}, \alpha_{\text{ub}}]$, we can show that $\beta_{k, 3}, \beta_{k, 4}, \beta_{k, 5}$ satisfy Condition 2 Lemma~\ref{lemma:psiconverge} and thus equation~\eqref{eq:betaasp} also holds for $p=3\sim5$. Considering that \eqref{eq:betaasp} holds for $p=1\sim 5$, through the union bound~\eqref{eq:betaunion}, we show that statement (e) holds. 

\textbf{(f)} According to the convexity of function $g_0(\cdot)$ from equation~\eqref{eq:galphadiff} Lemma~\ref{eq:gconcave}, we have $|g_0(\alpha)-g_0(\alpha^\star)|\leq N_1|\alpha-\alpha^\star|$. Therefore, we can upper bound the expected value of $\frac{1-\mathbb{E}[e^{-2\theta D}]}{2\theta}g_0(\alpha_k)$ as follows:
\begin{align}
    &\mathbb{E}\left[\left|\frac{1-\mathbb{E}[e^{-2\theta D}]}{2\theta}g_0(\alpha_k)\right|\right]\nonumber\\
    \leq&\mathbb{E}\left[\left|\frac{1-\mathbb{E}[e^{-2\theta D}]}{2\theta}N_1(\alpha^\star-\alpha_k)\right|\right]\nonumber\\
    \leq&\mathcal{O}(1/\sqrt{k}). 
\end{align}
This verifies Condition 2 in Lemma~\ref{lemma:psiconverge} and therefore verifies statement (f). 

\section{Proof of Theorem \ref{thm:freq}}\label{pf:thmfreq}
Recall that the sampling debt queue $U_k$ evolves as
\begin{equation*}
	U_{k+1} = \left(U_k + \frac{1}{f_{\text{max}}} - L_k \right)^+.
\end{equation*}
According to \cite{SNO2010Neely}, in order to satisfy the sampling constraint, it is sufficient to prove that
\begin{equation*}
	\limsup_{K \to \infty} \frac{1}{K} \sum_{k=1}^K \mathbb{E}[U_k] < \infty.
\end{equation*}

Here we adopt the Lyapunov drift-plus-penalty method to prove the stability of $U_k$. Define the Lyapunov function as
\begin{equation}\label{eq:L_Uk}
	L(U_k) = \frac{1}{2}U_k^2,
\end{equation}
and the Lyapunov drift is defined by
\begin{equation}\label{eq:Delta_k}
	\Delta(U_k) = \mathbb{E}[L(U_{k+1})-L(U_k)|U_k].
\end{equation}

First we upper bound $U_{k+1}^2$:
\begin{align*}
	U_{k+1}^2 &= \left[\max\{U_k + \frac{1}{f_{\text{max}}} - L_k, 0\}  \right]^2 \\
	&\le \left(U_k + \frac{1}{f_{\text{max}}} - L_k\right)^2.
\end{align*}

Plugging the above inequality into \eqref{eq:L_Uk} yields
\begin{align*}
	&L(U_{k+1}) - L(U_k) \\
	\le& \frac{1}{2} \left[\left(U_k + \frac{1}{f_{\text{max}}} - L_k \right)^2 - U_k^2 \right] \\
	=& -U_k\left(L_k-\frac{1}{f_\text{max}}\right) + \frac{1}{2} \left( \frac{1}{f_{\text{max}}} - L_k\right)^2.
\end{align*}

Plugging the above equation into \eqref{eq:Delta_k} and then take the expectation on both sides of \eqref{eq:Delta_k} yields
\begin{align}\label{eq:Delta_upp}
	&\Delta(U_k)\nonumber\\
	\overset{(a)}{\le}& -U_k\mathbb{E}\left[\left.L_k-\frac{1}{f_\text{max}} \right|U_k \right]\nonumber \\
	&+ \frac{1}{2} \left(\frac{1}{f_\text{max}^2} + \mathbb{E}[D_k^2]+\mathbb{E}[W_k^2|U_k]+2\mathbb{E}[D_kW_k|U_k]\right) \nonumber\\
	\le&-U_k\mathbb{E}\left[\left.L_k-\frac{1}{f_\text{max}} \right|U_k \right]\nonumber\\
	&+ \frac{1}{2} \left(\frac{1}{f_\text{max}^2} + M_{\text{ub}} + \mathbb{E}[W_k^2|U_k]+2\mathbb{E}[D_kW_k|U_k]\right).
\end{align}
where (a) holds since $D_k$ is independent of $U_k$. Similar to the proof of Lemma \ref{lem:bound}, we can bound $\mathbb{E}[D_kW_k|U_k]$ and $\mathbb{E}[W_k^2|U_k]$ as 
\begin{align*}
	&\mathbb{E}[D_kW_k|U_k] \le  D_\text{ub} \frac{v(\eta)^2}{\sigma^2} e^{\frac{2\theta}{\sigma^2}v(\eta)^2} \\
	&\mathbb{E}[W_k^2|U_k] \le \frac{2v(\eta)^3}{\sigma^3} \sqrt{\frac{\pi}{\theta}} e^{\frac{3\theta}{\sigma^2}v(\eta)^2}.
\end{align*}

Therefore, we have
\begin{align*}
	\Delta(U_k)\le& -U_k\mathbb{E}\left[\left.W_k+D_k-\frac{1}{f_\text{max}} \right|U_k \right] \\
	&+ \frac{1}{2} \left(\frac{1}{f_\text{max}^2} + M_{\text{ub}} + \frac{2v(\eta)^3}{\sigma^3} \sqrt{\frac{\pi}{\theta}} e^{\frac{3\theta}{\sigma^2}v(\eta)^2}\right.\\
	&\left.+2D_\text{ub} \frac{v(\eta)^2}{\sigma^2} e^{\frac{2\theta}{\sigma^2}v(\eta)^2}\right).
\end{align*}

Now we upper bound the first term of the RHS of \eqref{eq:Delta_upp}. According to \eqref{eq:W_k^*}, the waiting time $W_k$ is the optimal solution to
\begin{equation}\label{eq:infW}
    \sup_{w} \mathbb{E}\left[O_{D_k+w}^2- \left(\alpha_k-\lambda_k\right)w|O_{D_k},D_k \right].
\end{equation}

For simplicity, we denote the historical information $O_{D_k}, D_k$ to be $\mathcal{M}_{k-1}$.

Let $W_\epsilon$ be the waiting time under policy $\pi_\epsilon$. According to \eqref{eq:infW}, we have
\begin{align*}
    &\mathbb{E}[O_{D_k + W_k}^2|\mathcal{M}_{k-1}] - \mathbb{E}[(\alpha_k - \lambda_k) W_k|\mathcal{M}_{k-1}] \\
    \ge& \mathbb{E}[O_{D_k + W_\epsilon}^2|\mathcal{M}_{k-1}] - \mathbb{E}[(\alpha_k - \lambda_k) W_\epsilon|\mathcal{M}_{k-1}].
\end{align*}

Adding $\frac{1}{V}U_k\left(D_k-\frac{1}{f_\text{max}}\right)$ on both sides yields
\begin{align*}
    &\mathbb{E}[O_{D_k + W_k}^2|\mathcal{M}_{k-1}] - \mathbb{E}[(\alpha_k - \frac{1}{V}U_k) W_k|\mathcal{M}_{k-1}] \\
    &+\frac{1}{V}U_k\left(D_k-\frac{1}{f_\text{max}}\right) \\
    \ge& \mathbb{E}[O_{D_k + W_\epsilon}^2|\mathcal{M}_{k-1}] - \mathbb{E}[(\alpha_k - \frac{1}{V}U_k) W_\epsilon|\mathcal{M}_{k-1}] \\
    &+\frac{1}{V}U_k\left(D_k-\frac{1}{f_\text{max}}\right).
\end{align*}

Rearranging the terms yields
\begin{align*}
    &-U_k \mathbb{E}\left[\left.D_k+W_k-\frac{1}{f_\text{max}} \right|\mathcal{M}_{k-1}\right] \\
    \le&-U_k \mathbb{E}\left[\left.D_k+W_\epsilon-\frac{1}{f_\text{max}} \right|\mathcal{M}_{k-1}\right] \\
    & - V\mathbb{E}[O_{D_k + W_\epsilon}^2-\alpha_k W_\epsilon |\mathcal{M}_{k-1}] \\
    & + V\mathbb{E}[O_{D_k + W_k}^2-\alpha_k W_k |\mathcal{M}_{k-1}] \\
    \overset{(a)}{\le}&-U_k \epsilon
    + V\mathbb{E}[O_{D_k + W_k}^2+\alpha_k W_\epsilon|\mathcal{M}_{k-1}] \\
    \overset{(b)}{\le}& -U_k \epsilon+ V\left(\frac{\sigma^2}{2\theta} + \alpha_{\text{ub}}W_{\text{ub}}\right),
\end{align*}
where (a) holds by Assumption \ref{assu:eps}; (b) holds by Lemma \ref{lem:bound} and $W_{\text{ub}}=\frac{1}{f_\text{max}}+D_{\text{ub}}$ for sufficiently small $\epsilon$.

Now we have
\begin{align*}
	\Delta(U_k) \le& -U_k\epsilon + V\left(\frac{\sigma^2}{2\theta} + \alpha_{\text{ub}}W_{\text{ub}}\right) \\
	&+ \frac{1}{2} \left(\frac{1}{f_\text{max}^2} + M_{\text{ub}} + \frac{2v(\eta)^3}{\sigma^3} \sqrt{\frac{\pi}{\theta}} e^{\frac{3\theta}{\sigma^2}v(\eta)^2}\right.\\
	&\left.+2D_\text{ub} \frac{v(\eta)^2}{\sigma^2} e^{\frac{2\theta}{\sigma^2}v(\eta)^2}\right) \\
	\triangleq& -U_k\epsilon + C_1,
\end{align*}
where
\begin{align*}
	C_1 =&V\left(\frac{\sigma^2}{2\theta} + \alpha_{\text{ub}}W_{\text{ub}}\right) \\
	&+ \frac{1}{2} \left(\frac{1}{f_\text{max}^2} + M_{\text{ub}} + \frac{2v(\eta)^3}{\sigma^3} \sqrt{\frac{\pi}{\theta}} e^{\frac{3\theta}{\sigma^2}v(\eta)^2}\right.\\
	&\left.+2D_\text{ub} \frac{v(\eta)^2}{\sigma^2} e^{\frac{2\theta}{\sigma^2}v(\eta)^2}\right)
\end{align*}
is a constant. Summing up from $k=1$ to $K$ yields
\begin{align*}
	\mathbb{E}\left[\frac{1}{2} U_{k+1}^2 - \frac{1}{2} U_1^2\right] \le -\epsilon \sum_{k=1}^K \mathbb{E}[U_k] +KC_1.
\end{align*}

Notice that $ U_1 = 0$ and $U_{k+1} \ge 0$. Thus we have
\begin{equation*}
	\epsilon \sum_{k=1}^K \mathbb{E}[U_k] \le KC_1.
\end{equation*}

Rearranging the terms yields
\begin{equation*}
	\limsup_{K \to \infty} \frac{1}{K} \sum_{k=1}^K \mathbb{E}[U_k] \le \frac{C_1}{\epsilon} < \infty.
\end{equation*}

\section{Proof of Auxiliary Lemmas and Corollaries}\label{pf:aux}
\subsection{Proof of Corollary~\ref{cor:lbound}}\label{pf:lbound}
\begin{IEEEproof}
    \begin{align}
        &\left|l(\alpha)-l(\alpha^\star)\right|\nonumber\\
        =&\left|\mathbb{E}[D_k]+\mathbb{E}[\max\{R_1(v(\alpha))-R_1(|O_{D_k}|), 0\}]\right.\nonumber\\
        &\left.-\left(\mathbb{E}[D_k]+\mathbb{E}[\max\{R_1(v(\alpha^\star))-R_1(|O_{D_k}|), 0\}]\right)\right|\nonumber\\
        \leq&|R_1(v(\alpha))-R_1(v(\alpha^\star))|\nonumber\\
        \leq&\max_{\alpha\in[\alpha_{\text{lb}}, \alpha_{\text{ub}}]}\left|R_1'(v(\alpha))v'(\alpha)\right|\times\left|\alpha_k-\alpha^\star\right|. 
    \end{align}
\end{IEEEproof}

\subsection{Proof of Lemma \ref{lem:bound}}\label{pf:bound}

Since $O_{L_k}^2$ is an instance of $O_{D_k+W_k}^2$, we just bound $\mathbb{E}[O_{D_k+W_k}^2]$ and $\mathbb{E}[O_{D_k+W_k}^4]$. Therefore we have
\begin{align}
	&\mathbb{E}[O_{D_k+W_k}^2] = \frac{\sigma^2}{2\theta} \mathbb{E}[1 - e^{-2\theta(D_k + W_k)}] \le \frac{\sigma^2}{2\theta}. \\
	&\mathbb{E}[O_{D_k+W_k}^4] = 3 \mathbb{E}\left[\left(\frac{\sigma^2}{2\theta} (1 - e^{-2\theta(D_k + W_k)}) \right)^2 \right] \le \frac{3\sigma^4}{4\theta^2},\label{eq:OL4ub}
\end{align}
which verifies \eqref{eq:Qk_ub} and \eqref{eq:Qk2_ub}.
	
For $L_k$, according to Lemma \ref{lem:R1(v)} we can bound
\begin{align*}
	\mathbb{E}[L_k] &= \mathbb{E}[D_k] + \mathbb{E}[\max\{R_1(v(\alpha_k )) - R_1(|O_{D_k}|),0 \}]\\ 
	&\le D_\text{ub} + \mathbb{E}[R_1(v(\alpha_k))]. 
\end{align*}

Since $v(\alpha_k)$ is decreasing function with respect to $\alpha_k$, $v(\alpha_k)$ can be bounded
\begin{equation}
	0 < v(\alpha_{\text{ub}}) \le v(\alpha_k) \le v(\alpha_{\text{lb}}) \overset{(a)}{<} \infty,
\end{equation}
where (a) holds by Lemma \ref{lem:alpha_lb}.
	
Next, we bound $R_1(v)$ as
\begin{align*}
	R_1(v) =& \frac{v^2}{\sigma^2} {_2F_2} \left(1,1;\frac{3}{2},2;\frac{\theta}{\sigma^2}v^2 \right) \\
	=&\frac{v^2}{\sigma^2} \sum_{n=0}^\infty \frac{2^n}{(n+1)(2n+1)!!} (\frac{\theta}{\sigma^2}v^2)^n \\
	\overset{(a)}{\le}& \frac{v^2}{\sigma^2} \sum_{n=0}^\infty \frac{1}{n!} (\frac{2\theta}{\sigma^2}v^2)^n \\
	=& \frac{v^2}{\sigma^2} e^{\frac{2\theta}{\sigma^2}v^2}.
\end{align*}
where (a) holds by $n!\le (2n+1)!!$. Then we have
\begin{equation}\label{eq:R1_upper}
	0 \le R_1(v(\alpha_k)) \le \frac{v(\alpha_{\text{lb}})^2}{\sigma^2} e^{\frac{2\theta}{\sigma^2}v(\alpha_{\text{lb}})^2}.
\end{equation}
	
Therefore, we can bound $\mathbb{E}[L_k]$ as
\begin{align*}
	0 \le \mathbb{E}[L_k] \le& D_\text{ub} + \frac{v(\alpha_{\text{lb}})^2}{\sigma^2} e^{\frac{2\theta}{\sigma^2}v(\alpha_{\text{lb}})^2},
\end{align*}
which verifies \eqref{eq:Lk_ub}.

Finally, we rewrite $\mathbb{E}[L_k^2]$ as
\begin{align}
	\mathbb{E}[L_k^2] =& \mathbb{E}[(D_k + W_k)^2] \nonumber\\
	=& \mathbb{E}[D_k^2] + 2\mathbb{E}[D_kW_k] + \mathbb{E}[W_k^2] \nonumber\\
	=& \mathbb{E}[D_k^2] + 2\mathbb{E}[D_k\mathbb{E}[W_k|D_k]] + \mathbb{E}[W_k^2] \nonumber\\
	\le& M_\text{ub} + 2\mathbb{E}[D_k\mathbb{E}[W_k|D_k]] + \mathbb{E}[W_k^2].\label{eq:termL_k^2}
\end{align}

Next we bound $\mathbb{E}[D_k\mathbb{E}[W_k|D_k]]$ and $\mathbb{E}[W_k^2]$ respectively.
\begin{align*}
	\mathbb{E}[W_k|D_k] \overset{(a)}{\le}& \mathbb{E}[W_k | D_k, O_{D_k} < v(\alpha_k)] \\
	\overset{(b)}{=}& \mathbb{E}[R_1(v(\alpha_k)) - R_1(|O_{D_k}|)| D_k, |O_{D_k}| < v(\alpha_k)] \\
	\le& \mathbb{E}[R_1(v(\alpha_k))| D_k, |O_{D_k}| < v(\alpha_k)]\\
	\le& \mathbb{E}[R_1(v(\alpha_\text{lb}))] \\
	\overset{(c)}{\le}& \frac{v(\alpha_{\text{lb}})^2}{\sigma^2} e^{\frac{2\theta}{\sigma^2}v(\alpha_{\text{lb}})^2},
\end{align*}
where (a) holds because $W_k = 0$ if $ |O_{D_k}| > v(\alpha_k)$; (b) holds by Lemma \ref{lem:R1(v)}; (c) holds by \eqref{eq:R1_upper}. Therefore, we have
\begin{align}
	\mathbb{E}[D_k\mathbb{E}[W_k|D_k]] \overset{(a)}{\le}& \mathbb{E}[D_k]\frac{v(\alpha_{\text{lb}})^2}{\sigma^2} e^{\frac{2\theta}{\sigma^2}v(\alpha_{\text{lb}})^2} \nonumber \\
	\le& D_\text{ub} \frac{v(\alpha_{\text{lb}})^2}{\sigma^2} e^{\frac{2\theta}{\sigma^2}v(\alpha_{\text{lb}})^2}. \label{eq:termD_kW_k}
\end{align}
where (a) holds because $\frac{v(\alpha_{\text{lb}})^2}{\sigma^2} e^{\frac{2\theta}{\sigma^2}v(\alpha_{\text{lb}})^2}$ is a constant.

Now we bound $\mathbb{E}[W_k^2]$ as
\begin{align*}
	\mathbb{E}[W_k^2] =& \mathbb{E}[\mathbb{E}[W_k^2|D_k,\alpha_k]] \\
	\overset{(a)}{\le}& \mathbb{E}[\mathbb{E}[W_k^2|D_k,\alpha_k, |O_{D_k}|<v(\alpha_k)]]
\end{align*}
where (a) holds because $W_k = 0$ if $ |O_{D_k}| > v(\alpha_k)$. Now we just need to bound $\mathbb{E}[W_k^2|D_k,\alpha_k, |O_{D_k}|<v(\alpha_k)]$. According to \eqref{eq:alg_Wk}, $W_k$ is the stopping time that an OU process exits a bounded set $[-v(\alpha_k),v(\alpha_k)]$ with the initial state $O_{D_k}$. Denote $t^{(1)}_v(x)$ and $t^{(2)}_v(x)$ to be the first and second moment of $W_k$ with initial state $x$ and bounded set $[-v,v]$. According to \cite[Theorem 6.1]{Darling1953firstpassge}, we have
\begin{equation*}
	\frac{\sigma^2}{2}\frac{\text{d}^2 t^{(2)}_v(x)}{\text{d} x^2} - \theta x \frac{\text{d} t^{(2)}_v(x)}{\text{d} x} = -2t^{(1)}_v(x),~x\in[-v,v]
\end{equation*}
where according to Lemma \ref{lem:R1(v)}
\begin{equation}\label{eq:t^1(x)}
	t^{(1)}_v(x) = R_1(v) - R_1(x).
\end{equation}

Let $s(x)=\frac{\text{d} t^{(2)}_v(x)}{\text{d} x}$, and we have
\begin{equation*}
	\frac{\sigma^2}{2} s'(x) - \theta x s(x) = -2t^{(1)}_v(x). 
\end{equation*}

Multiplying $\frac{2}{\sigma^2} e^{-\frac{\theta}{\sigma^2}x^2}$ on both sides yields
\begin{equation*}
	s'(x) e^{-\frac{\theta}{\sigma^2}x^2} - \frac{2\theta}{\sigma^2} x s(x)e^{-\frac{\theta}{\sigma^2}x^2} = \frac{-4t^{(1)}_v(x)}{\sigma^2} e^{-\frac{\theta}{\sigma^2}x^2}.
\end{equation*}

This is equivalent to
\begin{equation*}
	\left(s(x)e^{-\frac{\theta}{\sigma^2}x^2} \right)' = \frac{-4t^{(1)}_v(x)}{\sigma^2} e^{-\frac{\theta}{\sigma^2}x^2}.
\end{equation*}

Therefore, we have
\begin{equation*}
	s(x) = Ce^{\frac{\theta}{\sigma^2}x^2} - e^{\frac{\theta}{\sigma^2}x^2}\int_{-v}^x \frac{4t^{(1)}_v(u)}{\sigma^2} e^{-\frac{\theta}{\sigma^2}u^2} \text{d}u,
\end{equation*}
where $C$ is a constant. Since $t^{(2)}_v(x)$ is even and takes the maximum when $x=0$. Therefore, we have
\begin{equation}\label{eq:C}
	C = \int_{-v}^0 \frac{4t^{(1)}_v(u)}{\sigma^2} e^{-\frac{\theta}{\sigma^2}u^2} \text{d}u.
\end{equation}

Since $t^{(2)}_v(x)$ is even, we only need to consider $x\in [-v,0]$. When $x\in [-v,0]$, $t^{(2)}_v(x)$ is increasing and $s(x) \ge 0$. Therefore, we have
\begin{equation}\label{eq:s(x)}
	0 \le s(x) \le Ce^{\frac{\theta}{\sigma^2}x^2}, ~x\in [-v,0].
\end{equation}

Then for $x\in [-v,0]$
\begin{align*}
	t^{(2)}_v(x) =& \int_{-v}^x s(u) \text{d}u \\
	\overset{(a)}{\le}& \int_{-v}^x Ce^{\frac{\theta}{\sigma^2}x^2} \text{d}u \\
	\le& \int_{-v}^0 Ce^{\frac{\theta}{\sigma^2}x^2} \text{d}u \\
	\overset{(b)}{=}& ve^{\frac{\theta}{\sigma^2}x^2} \int_{-v}^0 \frac{4t^{(1)}_v(u)}{\sigma^2} e^{-\frac{\theta}{\sigma^2}u^2} \text{d}u. \\
	\overset{(c)}{\le}& vR_1(v)e^{\frac{\theta}{\sigma^2}x^2} \int_{-v}^0 \frac{4}{\sigma^2} e^{-\frac{\theta}{\sigma^2}u^2} \text{d}u \\
	=& vR_1(v)e^{\frac{\theta}{\sigma^2}x^2} \frac{2}{\sigma} \sqrt{\frac{\pi}{\theta}} \text{erf}\left(\frac{\sqrt{\theta}}{\sigma}v \right) \\
	\overset{(d)}{\le}& vR_1(v)e^{\frac{\theta}{\sigma^2}v^2} \frac{2}{\sigma} \sqrt{\frac{\pi}{\theta}}.
\end{align*}
where (a) holds by \eqref{eq:s(x)}; (b) holds by \eqref{eq:C}; (c) holds by \eqref{eq:t^1(x)}; (d) holds since $\text{erf}(x)\le 1$. Since $t^{(2)}_v(x)$ is even for $x\in [-v,v]$, we have
\begin{equation*}
	t^{(2)}_v(x) \le \frac{2}{\sigma} \sqrt{\frac{\pi}{\theta}} vR_1(v)e^{\frac{\theta}{\sigma^2}v^2} ,~x\in [-v,v].
\end{equation*}

This means
\begin{align*}
	&\mathbb{E}[W_k^2|D_k,\alpha_k, |O_{D_k}|<v(\alpha_k)] \\
	\le& \frac{2}{\sigma} \sqrt{\frac{\pi}{\theta}} v(\alpha_k)R_1(v(\alpha_k))e^{\frac{\theta}{\sigma^2}v(\alpha_k)^2} \\
	\overset{(a)}{\le}& \frac{2}{\sigma} \sqrt{\frac{\pi}{\theta}} v(\alpha_\text{lb})\frac{v(\alpha_{\text{lb}})^2}{\sigma^2} e^{\frac{2\theta}{\sigma^2}v(\alpha_{\text{lb}})^2}e^{\frac{\theta}{\sigma^2}v(\alpha_\text{lb})^2} \\
	=& \frac{2v(\alpha_{\text{lb}})^3}{\sigma^3} \sqrt{\frac{\pi}{\theta}} e^{\frac{3\theta}{\sigma^2}v(\alpha_\text{lb})^2},
\end{align*}
where (a) holds by \eqref{eq:R1_upper}. Therefore we have
\begin{equation}\label{eq:W_k^2}
	\mathbb{E}[W_k^2] \le \frac{2v(\alpha_{\text{lb}})^3}{\sigma^3} \sqrt{\frac{\pi}{\theta}} e^{\frac{3\theta}{\sigma^2}v(\alpha_\text{lb})^2}.
\end{equation}

Plugging \eqref{eq:termD_kW_k} and \eqref{eq:W_k^2} into \eqref{eq:termL_k^2} yields
\begin{align*}
	\mathbb{E}[L_k^2] \le& M_\text{ub} + 2D_\text{ub} \frac{v(\alpha_{\text{lb}})^2}{\sigma^2} e^{\frac{2\theta}{\sigma^2}v(\alpha_{\text{lb}})^2}\\
	&+ \frac{2v(\alpha_{\text{lb}})^3}{\sigma^3} \sqrt{\frac{\pi}{\theta}} e^{\frac{3\theta}{\sigma^2}v(\alpha_\text{lb})^2},
\end{align*}
which verifies \eqref{eq:Lk2_ub}.
\subsection{Proof of Lemma~\ref{eq:gconcave}}\label{pf:gconcave}
\begin{IEEEproof}
For notational simplicity, for each stopping rule $w$, denote $\tilde{L}(w, \alpha-\lambda):=\mathbb{E}[-O_{D+w}^2+(\alpha-\lambda)w]$, which equals \eqref{eq:W_k^*}  before taking the infimum. Recall that the selection rule 
    \[w(O_D;\alpha-\lambda)=\inf\{t\geq D:|X_t-\hat{X}_t|\geq v(\alpha-\lambda)\}. \]
    is chosen to minimize function~\eqref{eq:W_k^*}. We have
    \begin{equation}
        -g_\lambda(\alpha)=\inf_w \tilde{L}(w, \alpha-\lambda). 
    \end{equation}
    For each policy $w$, function $\tilde{L}(w, \alpha-\lambda)$ is a linear increasing function of $\alpha$. Then by taking the infimum, function $\inf_w \tilde{L}(w, \alpha-\lambda)$ is continuous, concave and increasing. Therefore, function $g_\lambda(\alpha)$ is convex and monotonic decreasing. 

    When $\lambda^\star=0$, according to Lemma~\ref{lem:beta} equation \eqref{eq:opt_b}, we have $g_0(\alpha^\star)=0$. The derivative at $\alpha^\star$ can be computed by:
    \begin{equation}
        g_0'(\alpha^\star)=o'(\alpha^\star)-\alpha^\star l'(\alpha^\star)-l(\alpha^\star)\overset{(a)}{=}-l(\alpha^\star),
    \end{equation}
    where equation $(a)$ is obtained because $v(\alpha^\star)$ is the optimum threshold the minimizes $\mathbb{E}[-O_{D+w}^2+\alpha^\star (D+w)]$ so that $o'(\alpha)-\alpha^\star l'(\alpha)=0$. Then according to the convexity of $g_0(\cdot)$, the Taylor expansion at $\alpha^\star$ implies:
    \begin{align}
        g_0(\alpha)\geq&g_0(\alpha^\star)-l(\alpha^\star)(\alpha-\alpha^\star)\nonumber\\
        &+\frac{1}{2}\min_{\alpha'\in[\alpha_{\text{lb}}, \alpha_{\text{ub}}]}g_0''(\alpha')(\alpha-\alpha^\star)^2. 
    \end{align}
    
    Since function $g_0(\alpha)$ is monotically decreasing and convex, by taking $N=\frac{1}{2}\min_{\alpha'\in[\alpha_{\text{lb}}, \alpha_{\text{ub}}]}g_0''(\alpha')$, we have:
    \begin{equation}
        g_0(\alpha)\geq-l(\alpha^\star)(\alpha-\alpha^\star)+N(\alpha-\alpha^\star)^2. 
    \end{equation}

    From the convexity of $g_0(\cdot)$, we have:
    \begin{equation}
        g_0(\alpha)\geq g_0(\alpha^\star)-l(\alpha^\star)(\alpha-\alpha^\star). 
    \end{equation}
    
    Then notice that $g_0(\alpha)$ is monotonic decreasing, $g_0'(\alpha^\star)<0$, for $\alpha>\alpha^\star$, $g_0(\alpha)\leq 0$ and for $\alpha<\alpha^\star$, $g_0(\alpha)\geq 0$. Therefore we have
    \begin{equation}
        |g_0(\alpha)|\leq l(\alpha^\star)|\alpha-\alpha^\star|. 
    \end{equation}
\end{IEEEproof}

\begin{IEEEbiography}[{\includegraphics[width=1in,height=1.25in,clip,keepaspectratio]{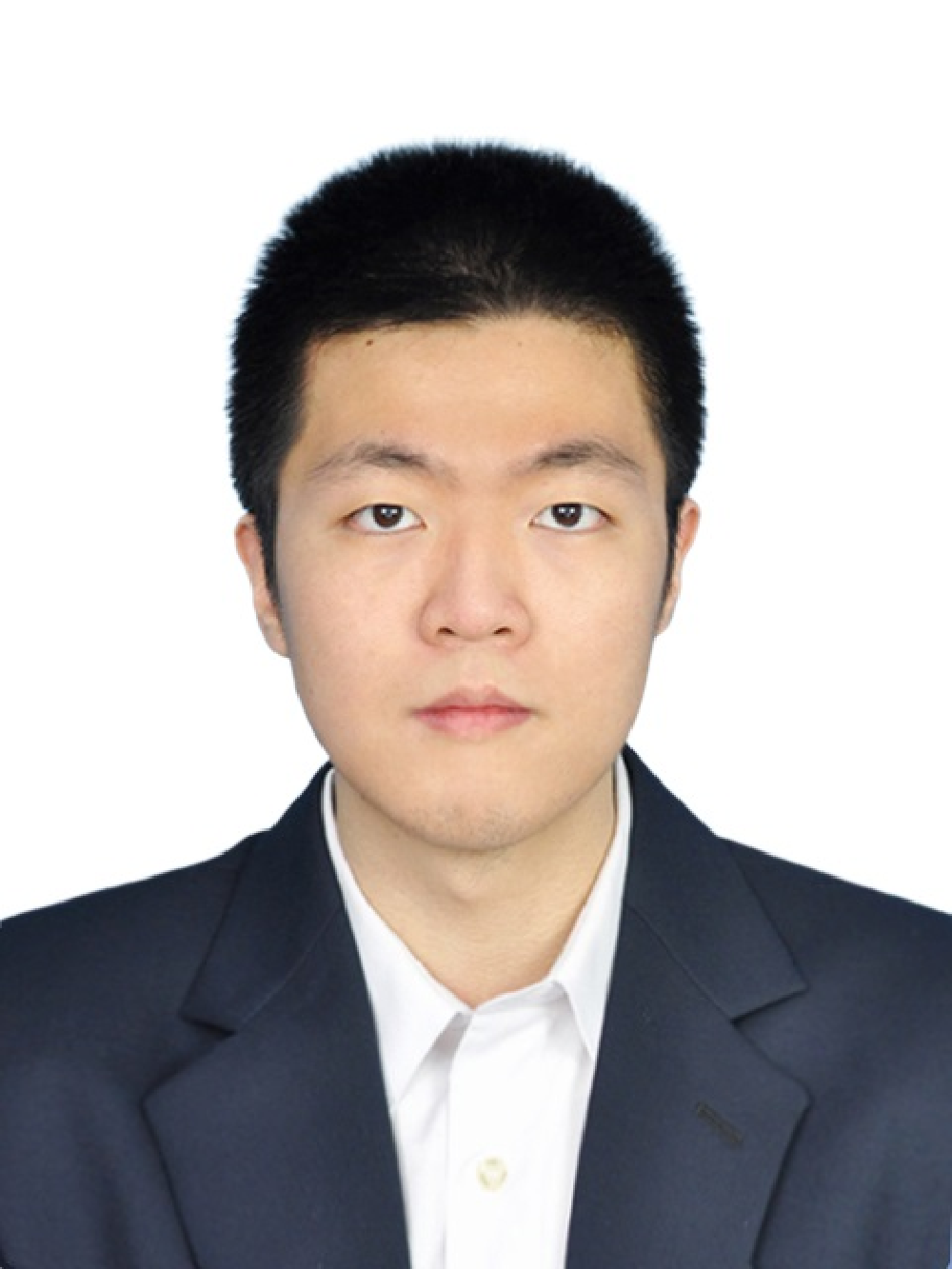}}]{Yuchao Chen}
	received the B.Eng. degree in electrical engineering from Tsinghua University, Beijing, China, in 2020. He is currently pursuing a Ph.D. degree at the Department of Electronic Engineering, Tsinghua University. His research interests include stochastic networking optimization, online learning, and wireless scheduling.
\end{IEEEbiography}

\begin{IEEEbiography}[{\includegraphics[width=1in,height=1.25in,clip,keepaspectratio]{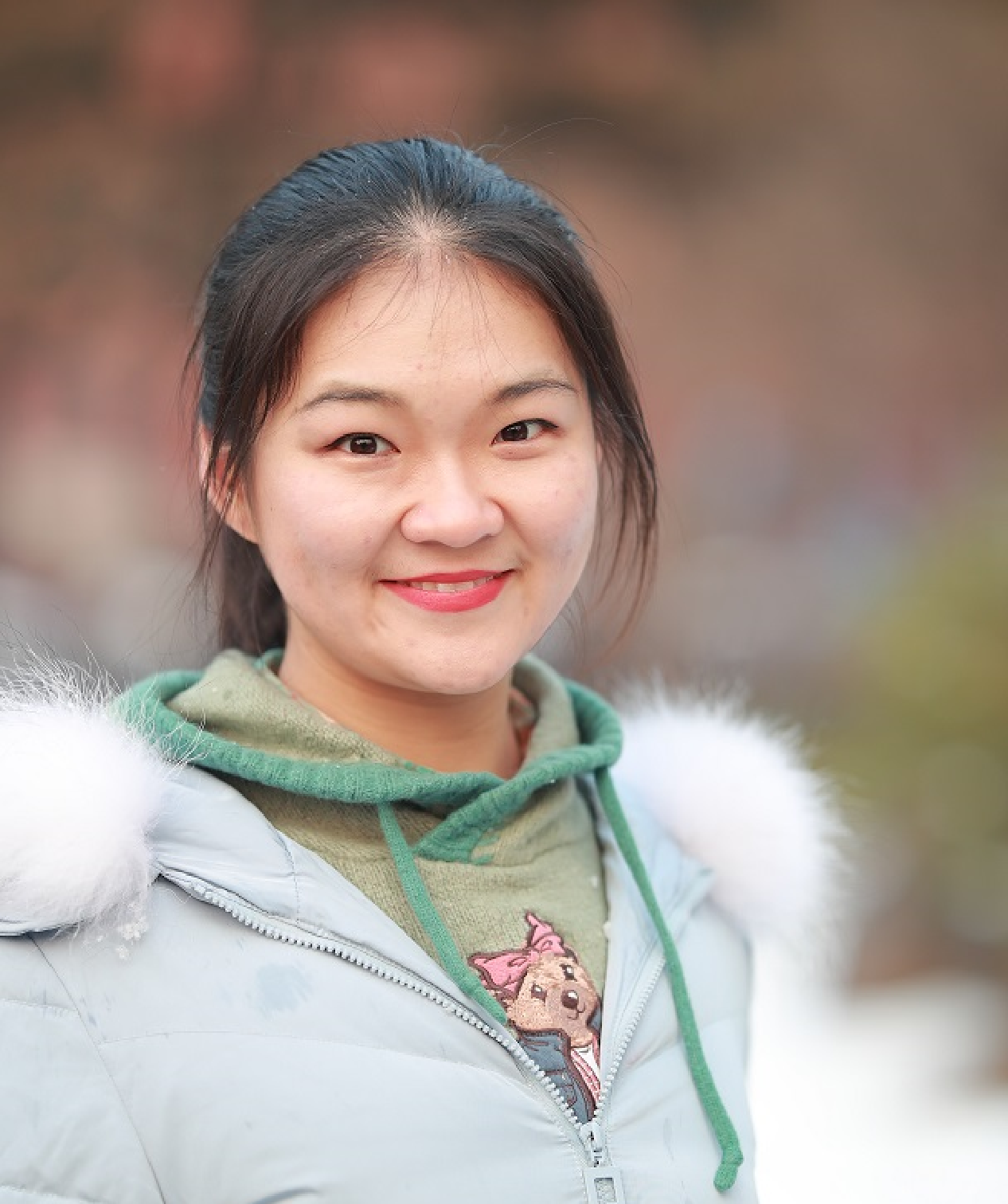}}]{Haoyue Tang}
Haoyue Tang (Student Member, IEEE) received the B.Eng. and Ph.D. degrees from the Department of Electronic Engineering, Tsinghua University, Beijing, China, in 2017 and 2022, respectively. She was a Postdoctoral Research Associate at Yale University from 2022-2023. She is currently a Post-Doctoral Research Associate at Meta AI. She was a selected participant at 2022 EECS Rising Stars workshop. Her research interests include age of information, stochastic network optimization, and statistical learning theory.
\end{IEEEbiography}

\begin{IEEEbiography}[{\includegraphics[width=1in,height=1.25in,clip,keepaspectratio]{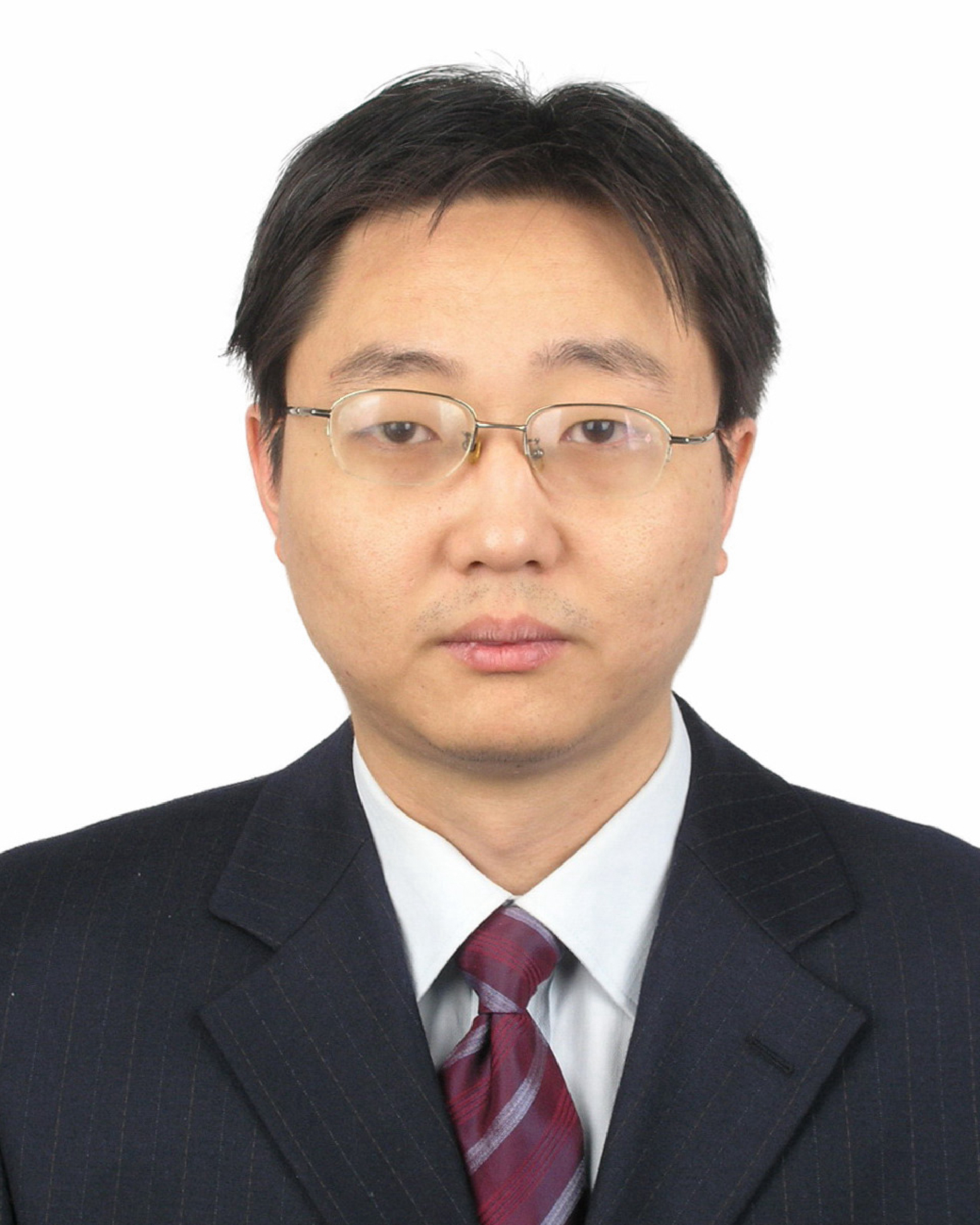}}]{Jintao Wang}
	(SM'12) received the B.Eng. and Ph.D. degrees in electrical engineering both from Tsinghua University, Beijing, China, in 2001 and 2006, respectively. From 2006 to 2009, he was an Assistant Professor in the Department of Electronic Engineering at Tsinghua University. Since 2009, he has been an Associate Professor and Ph.D. Supervisor. He is the Standard Committee Member for the Chinese national digital terrestrial television broadcasting standard. His current research interests include space-time coding, MIMO, and OFDM systems. He has published more than 100 journal and conference papers and holds more than 40 national invention patents.
\end{IEEEbiography}

\begin{IEEEbiography}[{\includegraphics[width=1in,height=1.25in,clip,keepaspectratio]{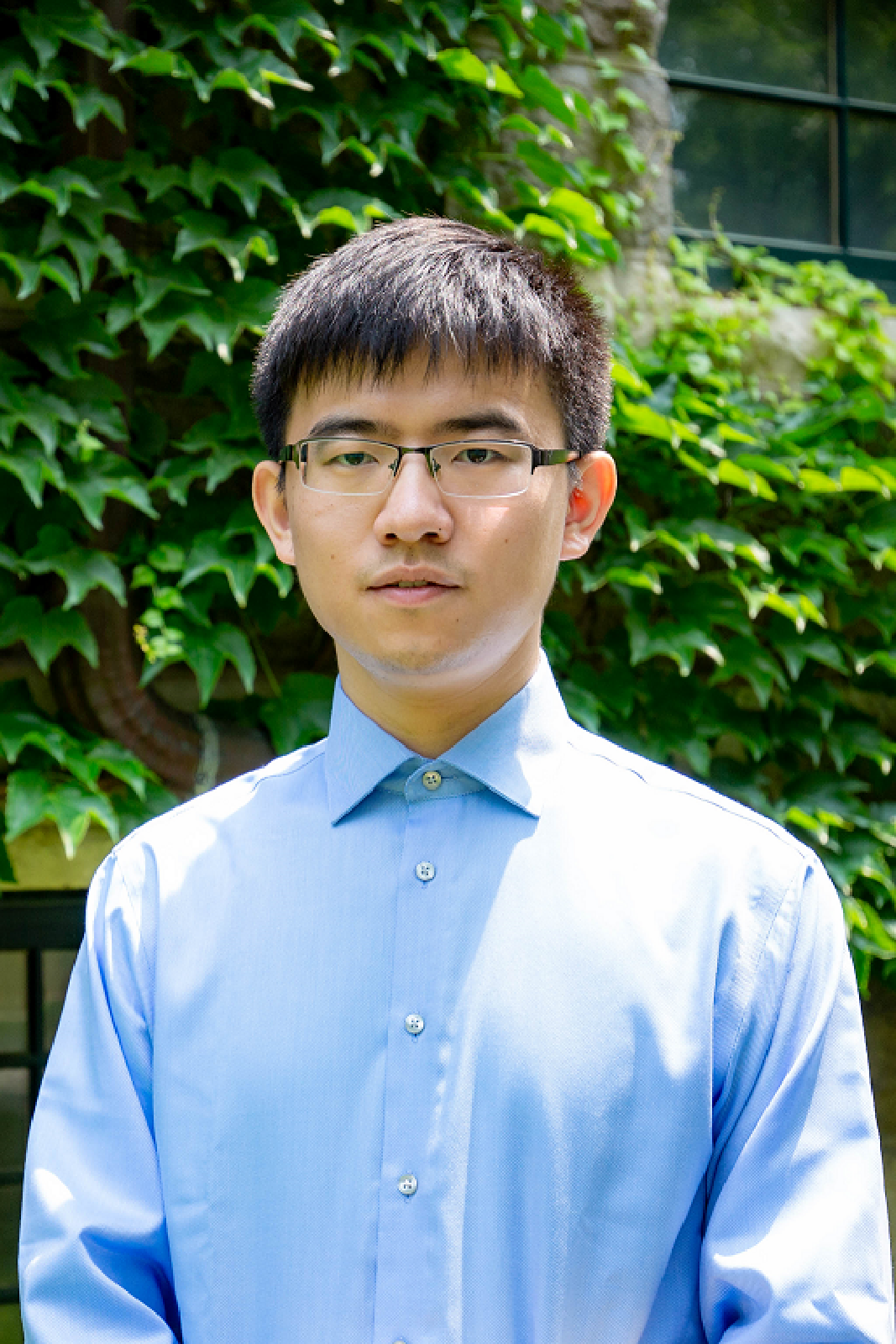}}]{Pengkun Yang}
	received the B.E. degree from the Department of Electronic Engineering, Tsinghua University, in 2013, the M.S. and Ph.D. degrees from the Department of Electrical and Computer Engineering, University of Illinois at Urbana–Champaign. He is currently an Assistant Professor with the Center for Statistical Science, Tsinghua University. His research interests include statistical inference, learning, and optimization and systems. He was a recipient of the Jack Keil Wolf ISIT Student Paper Award from the 2015 IEEE International Symposium on Information Theory.
\end{IEEEbiography}

\begin{IEEEbiography}[{\includegraphics[width=1in,height=1.25in,clip,keepaspectratio]{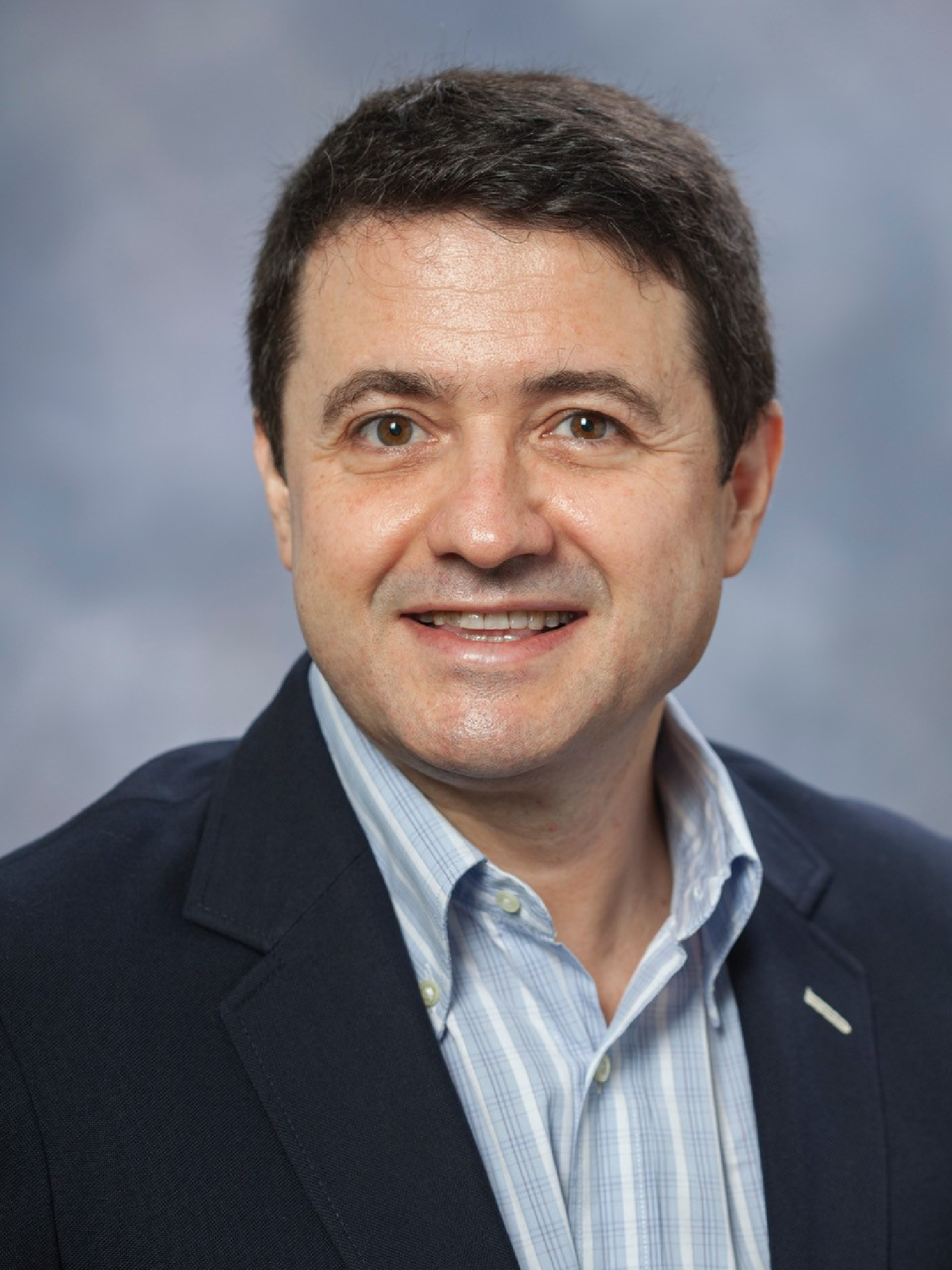}}]
{Leandros Tassiulas}(Fellow, IEEE) received the Ph.D. degree in electrical engineering from the University of Maryland, College Park, MD, USA, in 1991, and the Diploma degree in electrical engineering from the Aristotele University of Thessaloniki, Greece. He was a Faculty Member at the Polytechnic University, New York, NY, USA, University of Maryland, and University of Thessaly, Greece. He is currently the John C. Malone Professor of electrical engineering with Yale University, New Haven, CT, USA. His most notable contributions include the max-weight scheduling algorithm and the back-pressure network control policy, opportunistic scheduling in wireless, the maximum lifetime approach for wireless network energy management, and the consideration of joint access control and antenna transmission management in multiple antenna wireless systems. He was worked in the field of computer and communication networks with emphasis on fundamental mathematical models and algorithms of complex networks, wireless systems and sensor networks. His current research interests include intelligent services and architectures at the edge of next generation networks including the Internet of Things, sensing and actuation in terrestrial, and non terrestrial environments. His research has been recognized by several awards, including the IEEE Koji Kobayashi Computer and Communications Award in 2016, the ACM SIGMETRICS achievement award 2020, the Inaugural INFOCOM 2007 Achievement Award for fundamental contributions to resource allocation in communication networks, the INFOCOM 1994 and 2017 Best Paper Awards, the National Science Foundation (NSF) Research Initiation Award in 1992, the NSF CAREER Award in 1995, the Office of Naval Research Young Investigator Award in 1997, and the Bodossaki Foundation Award in 1999. He is a several best paper awards including the INFOCOM 1994, 2017 and Mobihoc 2016. He is a Fellow of ACM in 2020.
\end{IEEEbiography}
\end{document}